\documentclass{ifacconf}

\usepackage{graphicx}      
\usepackage{natbib}        

\usepackage{amsmath,amssymb,amsfonts}
\usepackage{graphicx}
\usepackage{textcomp}
\usepackage{xcolor}
\usepackage{cases}
\usepackage{mathrsfs}
\usepackage{amscd}
\usepackage{graphicx}  
\usepackage{epstopdf}
\usepackage{subfigure}
\usepackage{diagbox}
\usepackage{multirow}
\usepackage{booktabs}
\usepackage{epstopdf}
\usepackage{bm}
\allowdisplaybreaks
\usepackage{booktabs}
\usepackage{multirow}
\usepackage{stfloats}

\newtheorem{remark}{Remark}[section]

\newtheorem{example}{Example}[section]

\newcommand{\C}{{\mathbb C}}
\newcommand{\T}{{\mathbb T}}

\newcommand{\col}{\operatorname{col}}

\usepackage[noend]{algpseudocode}

\begin{document}
\begin{frontmatter}

\title{Asymptotical Cooperative Cruise Fault Tolerant Control for Multiple High-speed Trains with State Constraints } 


\author[First]{Zhixin Zhang}  
\author[Second]{Zhiyong Chen}

\address[First]{School of Automation, Central South University\\ Changsha, Hunan 410083, China \\ (e-mail: zhixin.zhang@csu.edu.cn).}
\address[Second]{Corresponding author \\
School of Engineering, The University of Newcastle \\ Callaghan, NSW 2308, Australia \\
 (e-mail: zhiyong.chen@newcastle.edu.au)} 

	\begin{abstract}
This paper investigates the asymptotical cooperative cruise fault tolerant control problem for multiple high-speed trains consisting of multiple carriages in the presence of actuator faults. A distributed state-fault observer utilizing the structural information of  faults  
is designed to achieve asymptotical estimation of states and faults of each carriage. The observer does not rely on choice of control input, and thus it is separated from controller design. Based on the estimated values of states and faults, a distributed fault tolerance controller is designed to realize asymptotical cooperative cruise control of trains under the dual constraints of ensuring both position difference and velocity difference of adjacent trains in specified ranges throughout the whole process.  	\end{abstract}



\end{frontmatter}

	\section{Introduction}
	
 High-speed trains (HSTs) have become an increasingly popular mode of travel worldwide due to their efficiency, reliability, and comfort; \cite{di2019ertms}, \cite{li2020impact}, \cite{cascetta2020economic}. Shortening train tracking interval to improve the existing railway utilization efficiency is an efficient and economic method to relieve the pressure of passenger transport. 
It was reported in \cite{dong2016cooperative,schumann2017increase,song2018development,felez2019model,cao2021tracking} that cooperative cruise control can realize synchronous operation of a large number of trains in a very short tracking distance, which is of great significance for improving railway utilization efficiency.
	
\begin{figure*}[t] 
		\centering 
		\includegraphics[width=0.8\textwidth]{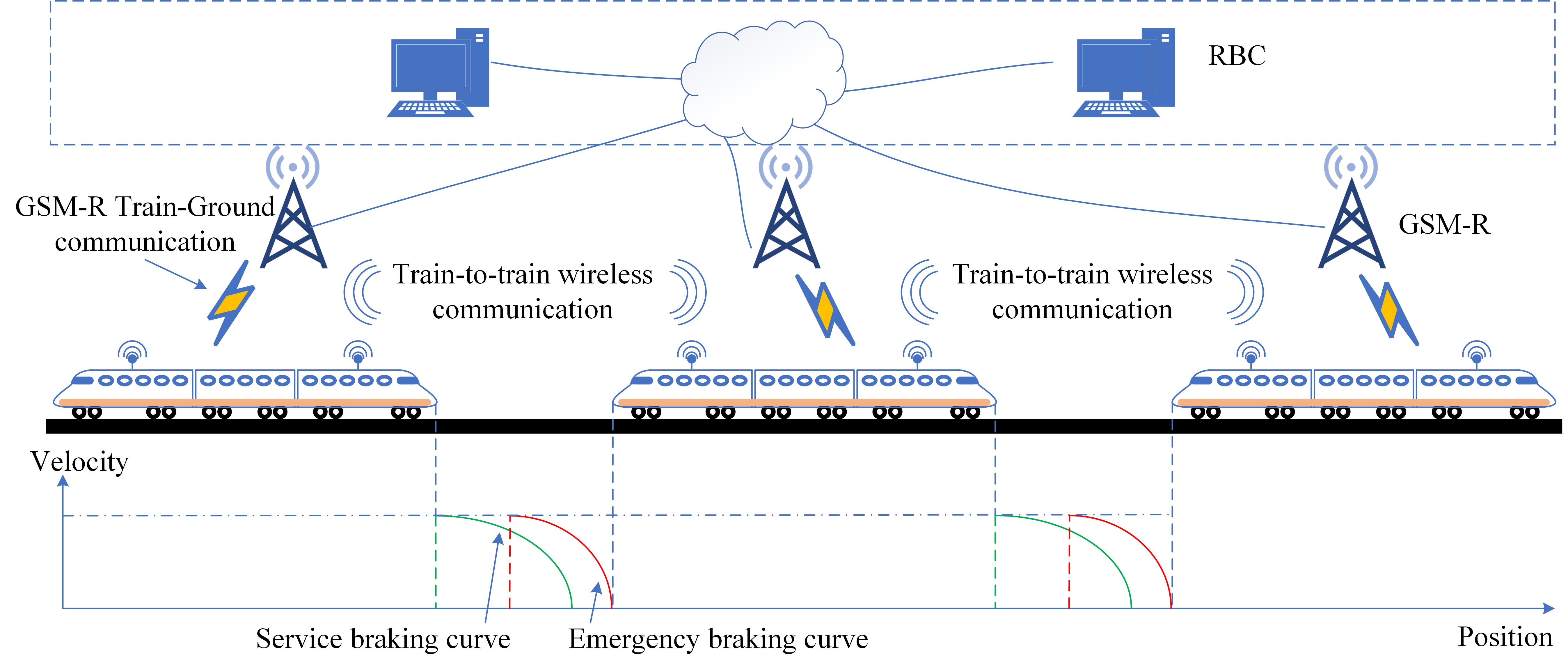} 
		\caption{Schematic diagram of cooperative operation of trains based on T2T wireless communication} 
		\label{fig_train_to_train} 
	\end{figure*}

The train-to-train (T2T) wireless communication technology was developed for HSTs 
in such as \cite{jikang2017switch, unterhuber2018path, song2019availability, wang2019robust,zhu2022distributed}, as shown in Fig.~\ref{fig_train_to_train}.  The global system for mobile communication-railway (GSM-R) network is responsible for bidirectional real-time information transmission between on-board equipment and the radio block center (RBC).	
This technology offers direct communication between adjacent trains and facilitates cooperative cruise control of trains.
	
A cooperative cruise control scheme has two tasks. On one hand, cruise control of a train aims to adjust the real-time velocity to track a reference velocity profile that is specified based on different  performance indexes such as energy consumption, comfort, and time consumption; see \cite{wang2013optimal, goverde2016three}. 
On the other hand, cooperative control of trains means that each train designs its controller based   on the real-time information transmitted by neighboring trains, and eventually achieves the same velocity on the premise of ensuring  safe operation of all trains; see \cite{dong2016cooperative, wang2019robust,  li2017cooperative}.

Cooperative cruise control of trains has attracted wide attention in recent years. 
The state-of-the-art research in this area is discussed from different perspectives below. 
For instance,  a single-particle train model was studied in \cite{li2015coordinated, gao2018cooperative, bai2020coordinated, dong2016cooperative,bai2021event, bai2021coordinated}, where the train is regarded as a rigid body and the torques 
between adjacent carriages are ignored. A more accurate  multiple-particle model treats every carriage as a mass point and considers the torque between carriages, which makes the design of the controller more flexible and realistic. 

In most references such as \cite{li2015coordinated, gao2018cooperative, wang2019robust, bai2020coordinated, dong2016cooperative, zhao2017distributed},  the train model is a second-order position-velocity model, which ignores the dynamic response from the designed input voltage of a motor to the real output torque, under an ideal assumption that the motor servo system is zero order. 
In practice, design of an ideal motor servo system is costly.
Therefore, it is of interest to study a more realistic model composed of a second-order position-velocity model 
and a dynamics motor servo system (treated as a first-order model), which results in a third-order position-velocity-acceleration
model. Some relevant work can be found in \cite{bai2021event,bai2021coordinated}. 
	
	
	It is a critical safety requirement to constrain the position spacing of trains in dynamic operation, as studied in literature such as \cite{li2015coordinated,bai2020coordinated, wang2019robust, zhao2017distributed}. 
	While these works only consider position constraint,  trains can be constrained at both the position and velocity levels to 
	 further improve the safety of operation.
	
	 Lastly, fault tolerance is critical for the cooperative cruise control scheme as
	potential actuator faults may threaten stable operation of trains and even cause serious accidents. 
There has been a lot of research on fault tolerance control of individual HSTs. For example, 
in \cite{mao2017adaptive},  an adaptive law was designed to update the parameters of the adaptive controller, so as to realize  fault compensation of HSTs under the condition that the system piecewise constant parameters and the actuator failure parameters are unknown.  In \cite{yao2018disturbance},   two Markov chains model were used to describe the failure process and
the failure detection and identification process of HSTs respectively, and an
 active fault tolerant composite hierarchical anti-disturbance
control strategy based on a disturbance observer was proposed.

In \cite{zhu2022distributed}, a distributed fault-tolerant control strategy was proposed to deal with the cooperative control problem of HSTs with actuator faults. However, the control strategy is aimed at the second-order single-particle train model, and does not consider the safety constraint of the distance between trains when multiple trains operate cooperatively, which brings safety hazards to short-distance tracking operation. Furthermore, there are few researches on 
cooperative control of the third-order train model  with actuator faults.
To the best of our knowledge, the research on fault tolerance control of the third-order train model was limited to the problem of a single train in \cite{song2013fault}.

Most studies on fault-tolerant control assume that the fault model is unknown but additional bounded constraints or other conditions need to be added.  These methods always lead to bounded error of fault estimation and thus bounded error of fault tolerance control; see
\cite{ge2019rbfnn, mao2019adaptive, lin2018adaptive, yao2018disturbance,  zhu2022distributed}.
However,  actual train actuator faults are mainly constant faults and periodic faults as seen in \cite{mao2019adaptive,liu2019adaptive}.  By utilizing such structural information,  a fault estimation observer was proposed 
in \cite{zhang2023fault} for a class of second-order systems, and a fault tolerance controller was designed based on the 
observed states to achieve asymptotic tracking. The approach is further developed in this paper 
for a third-order system and applied to HSTs to realize  cooperative cruise fault-tolerant control.
 
 The main contributions are summarized as follows. 
	
	\begin{itemize}
		\item[1.] A class of third-order multi-particle model of multiple HSTs with actuator faults is established. The fault is described based on a new representation, which contains the structural information and is of great significance for 
		the asymptotical estimation.
		
		\item[2.] A distributed state-fault observer is designed for every carriage of every train, which can realize asymptotical estimation of states and faults.  The observer does not rely on the control input, which separates designs of observer and controller
		and provides great convenience in theoretical analysis.
		
		\item[3.] A distributed cooperative cruise fault-tolerant controller is designed for all carriages of all trains.  
		All carriages of each train can asymptotically track the velocity profile of the head carriage, and  the position difference between every adjacent carriages of a same train converges to the nominal value. Also, the position difference and velocity difference of adjacent trains are constrained within the specified range  throughout the whole process,  the velocity of every train aymptotically tracks the reference velocity,  and the distance between adjacent trains eventually converges to the prescribed distance.
		
	\end{itemize}
		
	The rest of this paper is organized as follows. The dynamic model and the control objective are given in Section~\ref{sec_modeling}. Section~\ref{sec_observer} presents the design of a state-fault observer. In Section~\ref{sec_controller}, a cooperative cruise fault tolerance controller is explicitly developed. The numerical simulation results are presented in Section~\ref{sec_simulation}. Finally,  Section~\ref{sec_conclusion} concludes this paper.
	
	\section{Problem Formulation and Preliminaries} \label{sec_modeling}
	
	\subsection{Modeling}

Consider a railway network system of $N\geq 1$ trains, labelled as $i=1,\cdots, N$, 
each of which is composed of $M_i\geq 1$ carriages, labelled as $j=1,\cdots, M_i$.
Denote the sets of trains and carriages as 
	\begin{align*}
\T = & \{ i \; | \; i = 1,\cdots,N \}, \\
\C = & \{ (i,j) \; | \;  j= 1,\cdots,M_i,\; i\in \T \},
	\end{align*}
respectively.  
Also, denote the sets of head carriages and other (non-head) carriages 
as
\begin{align*}
	\C_h =& \{(i,j) \ |\  j = 1,\;i\in \T  \},\\
	\C_o =& \{ (i,j)\ |\   j = 2, \cdots, M_i ,\; i\in \T \}, 
	\end{align*}
respectively. One has $\C = \C_h \cup \C_o$.

The dynamics of each carriage are described as follows: 
	\begin{align}
		\dot{x}_{ij} =& v_{ij} \nonumber\\
		\dot{v}_{ij} =& \frac{1}{m_{ij}}(\tau_{ij} - B_{ij}(x_i, v_i) - m_{ij}R (v_{ij})),\; (i,j)\in \C,  \label{eq_sys_1}
	\end{align}
where $x_{ij}$, $v_{ij}$, and $\tau_{ij}$ are the position,  the velocity, 
and the input  representing the traction/braking force, respectively.  
The mass of carriage is $m_{ij}$, the function $R(v_{ij})$ characterizes the
aerodynamic drag and rolling mechanical resistance, and the term
$B_{ij}(x_i, v_i)$, with the lumped vectors $x_i=[x_{i1},\cdots, x_{i,M_i}]^T$
and  $v_i=[v_{i1},\cdots, v_{i,M_i}]^T$, represents the coupling force caused by the adjacent carriage(s).
Specifically, \begin{align*}
		R (v_{ij})= c_{0} + c_{1}v_{ij} + c_{2}v_{ij}^2 ,
	\end{align*}
where  $c_{0}$, $c_{1}$ and $c_{2}$ are the Davis formula coefficients; and
 \begin{align*}
		B_{ij}(x_i, v_i) = \begin{cases}
		&a(x_{i1} - x_{i2} - d_p) + b(v_{i1} - v_{i2})  ,\  j = 1 \vspace{2mm} 	\\
		&a(2x_{ij} - x_{i(j-1)} - x_{i(j+1)}) + b(2v_{ij} \\
		& - v_{i(j-1)} - v_{i(j+1)})  ,\  j = 2,\cdots,M_i-1 \vspace{2mm}  \\
		&a(x_{iM_i} - x_{i(M_i-1)} + d_p) + b(v_{iM_i} \\
		& - v_{i(M_i-1)})  ,\  j = M_i 		\end{cases},
	\end{align*}
where $d_p$, $a$ and $b$  are the nominal length (including the length of a carriage), the stiffness coefficient and the damping constant of carriage couplers, respectively.  



 Assume the dynamic response of the actuator for generating  $\tau_{ij}$ is  characterized by the following first-order  differential equation
	\begin{align}
		\dot{\tau}_{ij} = -r_{ij}\tau_{ij} + \varpi_{ij} + E_{ij}f_{ij},\; (i,j)\in \C  \label{eq_tau_d_2}
	\end{align}
where $\varpi_{ij}$ is the desired  force to be designed, 
$r_{ij}$ is the time constant, and  $E_{ij}f_{ij}$ represents the fault that may occur in the actuator. 
 The servomechanism for generating such closed-loop regulation performance is out of the scope of this paper. 

Assume the  fault $f_{ij}$ has the model  	
	\begin{align}
\dot{f}_{ij} =& S_{ij}f_{ij} ,\; (i,j)\in \C \label{eq_tau_d_1}
\end{align}
for a constant matrix $S_{ij}$.
Train actuator faults are 
typically  constant faults and periodic faults as studied in \cite{mao2019adaptive, liu2019adaptive}.
They can be well accommodated by the model \eqref{eq_tau_d_1}.
It is worth mentioning that most of the existing works neglect the fault model, which leads to 
residual fault estimation errors and thus affects accuracy of fault tolerance control. In this paper, a reasonable model is established for the common constant fault and periodic fault, and the fault model information is effectively exploited, which is beneficial to asymptotical estimation of the faults and  improvement of control accuracy.

\begin{example} \label{exa_fault} Consider the fault  $f_{ij}$ of the model  \eqref{eq_tau_d_1} with 
		\begin{align*}
f_{ij} =\begin{bmatrix} f_{ij1} \\ f_{ij2} \\ f_{ij3} \end{bmatrix} ,\;  S_{ij} =  \begin{bmatrix} 0 & 0 & 0\\ 0 & 0 & \omega_{ij}\\ 0 & -\omega_{ij} & 0 \end{bmatrix} .
	\end{align*}
Then, the  fault  $f_{ij}$ contains a constant mode and a sinusoidal model of frequency $\omega_{ij}$.
 In particular, if we select $E_{ij} =  \begin{bmatrix} \upsilon_{ij} & 0 & \nu_{ij}\omega_{ij} \end{bmatrix}$, the fault
 in \eqref{eq_tau_d_2} is 
 	\begin{align*}
 E_{ij}f_{ij} = \upsilon_{ij}f_{ij1} +    \nu_{ij}\omega_{ij}  f_{ij3}.  \end{align*}
 Assume the faults occur in certain periods,  one has
    	\begin{align*}
	f_{ij1}(t) =&\left\{ \begin{array}{ll} {F_{cij}}, & t\in [t^s_{ij1}, t^e_{ij1}] \\
	0 , &\text{otherwise} \end{array} \right. \\
	f_{ij3}(t) =&\left\{ \begin{array}{ll} {F_{pij}}\cos (\omega_{ij}t + {F_{\phi ij}}), & t\in [t^s_{ij3}, t^e_{ij3}] \\
	0 , &\text{otherwise} \end{array} \right.
    \end{align*}
 for some unknown parameters ${F_{cij}}$, ${F_{pij}}$, ${F_{\phi ij}}$. 
%
\end{example}

 \subsection{Preliminary Manipulation} \label{sec_prelim}
	
Let $w_{ij} := \dot{v}_{ij}$ be the acceleration of the carriage $(i,j)$, which is not measurable 
in the present setting due to practical restriction. 
Taking $w_{ij}$ as a new state, we can calculate $\dot w_{ij} = \ddot{v}_{ij}$ as follows,  
	\begin{align}   
		\dot w_{ij}
		=& -\frac{r_{ij}}{m_{ij}}\tau_{ij} + \frac{\varpi_{ij}}{m_{ij}} + \frac{E_{ij}}{m_{ij}}f_{ij} \nonumber \\
		&-\frac{1}{m_{ij}}\frac{d B_{ij}(x_i, v_i) }{dt}-   \frac{\partial R(v_{ij})}{\partial v_{ij}} \dot v_{ij} \nonumber \\
		=& -r_{ij} w_{ij} -\frac{r_{ij}}{m_{ij}}  B_{ij}(x_i, v_i)  -r_{ij}  R(v_{ij}) +\frac{\varpi_{ij}}{m_{ij}} \nonumber \\
		&  + C_{ij} f_{ij} -\frac{1}{m_{ij}} \frac{d B_{ij}(x_i, v_i) }{dt}-  \frac{\partial R(v_{ij})}{\partial v_{ij}} w_{ij}   \nonumber\\
		=& -\frac{r_{ij}}{m_{ij}}  B_{ij}(x_i, v_i)  - r_{ij}R (v_{ij}) + \frac{\varpi_{ij}}{m_{ij}} + C_{ij}f_{ij} \nonumber\\
		& + B^1_{ij}(v_{ij})w_{ij} + B^2_{ij}w_{i(j-1)} + B^3_{ij}w_{i(j+1)} + B^4_{ij}(v_i) \label{dwij}
	\end{align}
where $\tau_{ij}   = m_{ij}w_{ij} + B_{ij}(x_i, v_i)  + m_{ij}R(v_{ij})$, by \eqref{eq_sys_1}, is used, and the quantities 
are defined as follows,
\begin{align*}
	&C_{ij} = E_{ij}/m_{ij} \\
		&B^1_{ij}(v_{ij}) = \left\{ \begin{array}{l}
		 -\frac{b}{m_{ij}} - (c_{1} + 2c_{2}v_{ij}) - r_{ij}  ,\  j = 1,M_i \\
		 -\frac{2b}{m_{ij}} - (c_{1} + 2c_{2}v_{ij}) - r_{ij}  ,\  j = 2,\cdots,M_i-1 
		\end{array}\right.  \nonumber\\
		&B^2_{ij} =  \left\{ \begin{array}{l}
		 0   ,\  j = 1 \\
		 \frac{b}{m_{ij}}   ,\  j = 2,\cdots,M_i 
		\end{array}\right.  \nonumber\\
		&B^3_{ij} =  \left\{ \begin{array}{l}
		 \frac{b}{m_{ij}}   ,\  j = 1,\cdots,M_i-1 \\
		 0   ,\  j = M_i
		\end{array}\right.  \nonumber\\
		&B^4_{ij}(v_i) = \left\{ \begin{array}{l}
		-\frac{a(v_{i1} - v_{i2})}{m_{ij}}  ,\  j = 1 \\
		-\frac{a(2v_{ij} - v_{i(j-1)} - v_{i(j+1)})}{m_{ij}} ,\  j = 2,\cdots,M_i-1 \\
		-\frac{a(v_{iM_i} - v_{i(M_i-1)})}{m_{ij}} ,\  j = M_i 
		\end{array}\right. . 
	\end{align*}

The preliminary controller $\varpi_{ij}$ in \eqref{eq_tau_d_2} is constructed as follows, 
	\begin{align}
  \varpi_{ij} =& m_{ij} u_{ij}  + r_{ij}  B_{ij}(x_i, v_i) +m_{ij} r_{ij} R(v_{ij}) \nonumber \\ &  -m_{ij} B^4_{ij}(v_i) ,\; (i,j)\in \C ,
		\label{varpiij}
	\end{align}
where $u_{ij}$ is the new input to be designed.

From the above manipulation, the system composed of 
 \eqref{eq_sys_1}, \eqref{eq_tau_d_2}, \eqref{eq_tau_d_1}, \eqref{dwij}, and \eqref{varpiij}
is equivalent to the following composite model  
	\begin{align}
	\dot{x}_{ij} =& v_{ij} \nonumber\\
	\dot{v}_{ij} =& w_{ij} \nonumber\\
	\dot{w}_{ij} =&  B^1_{ij}(v_{ij})w_{ij} + B^2_{ij}w_{i(j-1)}  + B^3_{ij}w_{i(j+1)} \nonumber\\& + C_{ij}f_{ij} + u_{ij} \nonumber\\
	\dot{f}_{ij} =& S_{ij}f_{ij},\; (i,j)\in \C, \label{eq_sys_1_observed}
	\end{align}
which will be studied in the remaining sections of this paper. 
For the convenience of presentation, the full state of \eqref{eq_sys_1_observed} is denoted as 
 $\chi_{ij} = \col(x_{ij}, v_{ij}, w_{ij},f_{ij})$.

	\subsection{Objectives} \label{sec_objective}
	
Accurate regulation of trains to a velocity-position profile is of great significance to ensure safe running distance and punctual running time. Let the desired position, velocity, acceleration, and control input be $x_0$, $v_0$, $w_0$, and $u_0$ respectively, and they 
 satisfy
	\begin{align}
	\dot{x}_0 = v_0,\ \dot{v}_0 = w_0,\ \dot{w}_0 = u_0. \label{eq_ref_signal}
	\end{align}
For the system  \eqref{eq_sys_1_observed}, the objective of this paper is to design $u_{ij}$
such that the closed-loop system satisfies the requirements ${\bf R1}$, ${\bf R2}$, and ${\bf R3}$
defined below. 

 (i) For each train, the position difference between every adjacent carriages converges to the nominal value $d_p$,
 and the velocity difference converges to zero. Therefore, the first control requirement is
		\begin{align}
	{\bf R1:}	\lim_{t\rightarrow \infty} [x_{i(j-1)}(t) - x_{ij}(t)] =& d_p \nonumber\\
		\lim_{t\rightarrow \infty} [v_{i(j-1)}(t) - v_{ij}(t) ]=& 0 ,\; (i,j)\in \C_o. \label{eq_objective_1}
		\end{align}

(ii) To ensure reliable train-to-train communication, the distance between two adjacent trains must be less than the maximum communication radius $\gamma_1>0$. In addition, in order to avoid the risk of collision, 
the distance  must be larger than the emergency braking distance $\gamma_2>0$. 
 In order to improve utilization of tracks, it is also a requirement to ensure 
 that the distance is asymptotically maintained at the service braking distance $d_s$ satisfying 
 $\gamma_2 < d_s <\gamma_1$. Denote $\rho_1 = \gamma_1 - d_s$ and $\rho_2 = d_s - \gamma_2$.	
Let 
		\begin{align} 
		\tilde{x}_{i}(t) =\epsilon_i(t) - d_s ,\; \epsilon_i =& x_{(i-1)M_{i-1}} - x_{i1}, \; i \in\T \label{eq_ds}		\end{align}
where  $\epsilon_i$ is the distance between the $i$-th and $(i-1)$-th trains. 
	Then, the second control requirement is
		\begin{align}
		{\bf R2:}	&-\rho_2 < \tilde{x}_{i}(t) < \rho_1,\ \forall t \geq 0 \nonumber\\
			&\lim_{t\rightarrow \infty} \tilde{x}_{i}(t) = 0,\; i \in \T. \label{eq_objective_2}
		\end{align}
 It is noted that,  ${\bf R2}$ is equivalent to 
		\begin{align*}
			&\gamma_2 < \epsilon_i(t) < \gamma_1,\ \forall t \geq 0\\
			& \lim_{t\rightarrow \infty} \epsilon_i(t) = d_s,\; i \in \T.
		\end{align*}

(iii) In order to achieve stable communication between trains and reduce the risk of train collision, it is of great significance to restrict the velocity difference between adjacent trains within the prescribed range $(-\sigma_2, \sigma_1)$ and eventually converge to zero, with $\sigma_1>0$ and $\sigma_2>0$. Define $\tilde{v}_{i}(t) = v_{(i-1)M_{i-1}} - v_{i1}$, then 
		\begin{align}
			{\bf R3:}		&-\sigma_2 < \tilde{v}_{i}(t) < \sigma_1,\ \forall t\geq 0 \nonumber\\
			&\lim_{t\rightarrow \infty} \tilde{v}_{i}(t) = 0,\; i \in \T. \label{eq_objective_3}
		\end{align}

\begin{remark} 
\label{remark_definition_q}

For complement of notation, it is assumed the  
the $0$-th train is a virtual train whose model is \eqref{eq_ref_signal} 
and $x_{(i-1)M_{i-1}} = x_0$ and $v_{(i-1)M_{i-1}} = v_0$ for $i = 1$.
	Select any $0<\ell_{i1} < \min\{\sigma_2/\rho_1, \sigma_1/\rho_2\}$. 
	Let  $\varrho_{1} = -\ell_{i1}\rho_2 + \sigma_1>0$ and $\varrho_{2} = -\ell_{i1}\rho_1 + \sigma_2>0$.
	 Define a new error combination term $\tilde{q}_{i}(t) = \tilde{v}_{i}(t) + \ell_{i1}\tilde{x}_{i}(t)$  and the associated requirement
		\begin{align}
		{\bf R3':} &-\varrho_{2} < \tilde{q}_{i}(t) < \varrho_{1},\ \forall t \geq 0 \nonumber\\
		&\lim_{t \rightarrow \infty} \tilde{q}_{i}(t) = 0,\; i \in \T . \label{eq_objective_3_change}
		\end{align}
When the requirements \eqref{eq_objective_2} and \eqref{eq_objective_3_change} are satisfied, 
so is the requirement  \eqref{eq_objective_3}. Therefore, ${\bf R3}$ can be replaced by ${\bf R3'}$.
	\end{remark}

%
	\begin{remark} \label{remark_measurable}
		The reference singles $x_0$, $v_0$, $w_0$, and $u_0$ in \eqref{eq_ref_signal} are known for the head carriage of the first train. The state variables $x_{ij}$ and $v_{ij}$ are measurable by
		means of a speedometer and a milemeter equipped on a train
		(or other modern advanced technology such as Global Positioning System, Doppler radar, microwave radar, etc.).
	\end{remark}

	
%

	\section{Observer Design} \label{sec_observer}

To address the challenge caused by the existence of unknown faults $f_{ij}$ and unmeasurable states
$w_{ij}$, we first design an observer in this section to achieve accurate estimation of $f_{ij}$ and $w_{ij}$, and then design 
a controller based on the estimated values in the next section to achieve 
the three control requirements. The observer is designed as follows, with the state $\hat\chi_{ij} = \col(\hat x_{ij}, \hat v_{ij}, \hat w_{ij},\hat f_{ij})$,
	\begin{align}
		\dot{\hat{x}}_{ij} =& \hat{v}_{ij} + \mu_{{ij}1} \nonumber\\
		\dot{\hat{v}}_{ij} =& \hat{w}_{ij} + \mu_{{ij}2} \nonumber\\
		\dot{\hat{w}}_{ij} =& B^1_{ij}(v_{ij})\hat{w}_{ij} + B^2_{ij}\hat{w}_{i(j-1)} \nonumber\\
		& + B^3_{ij}\hat{w}_{i(j+1)} + C_{ij}\hat{f}_{ij} + u_{ij} + \mu_{{ij}3} \nonumber\\
		\dot{\hat{f}}_{ij} =& S_{ij}\hat{f}_{ij} + \mu_{{ij}4},  (i,j)\in \C \label{eq_observer_1}
	\end{align}
	where $\hat{x}_{ij}$, $\hat{v}_{ij}$, $\hat{w}_{ij}$, and $\hat{f}_{ij}$   are the estimation of $x_{ij}$, $v_{ij}$, $w_{ij}$, and $f_{ij}$, respectively.
	The auxiliary inputs $\mu_{{ij}1}$, $\mu_{{ij}2}$, $\mu_{{ij}3}$ and $\mu_{{ij}4}$ can be designed as follows.
	\begin{align}
	\mu_{{ij}1} =& -k_{{ij}1}(\hat{x}_{ij} - x_{ij}) + (v_{ij} - \hat{v}_{ij}) \nonumber\\
	\mu_{ij2} =& D^1_{ij}(v_{ij}) - D^1_{ij}(\hat{v}_{ij}) + k_{ij2}(\hat{v}_{ij} - v_{ij}) \nonumber\\
		& + D^2_{ij}(v_{i(j-1)}) - D^2_{ij}(\hat{v}_{i(j-1)}) \nonumber\\
		& + D^3_{ij}(v_{i(j+1)}) - D^3_{ij}(\hat{v}_{i(j+1)}) \nonumber\\
	 \mu_{ij3} 	=& B^1_{ij}(v_{ij})\mu_{ij2} + k_{ij3}(\hat{v}_{ij} - v_{ij}) \nonumber\\
		&+ (B^1_{ij}(\hat{v}_{ij}) - B^1_{ij}(v_{ij}))(\hat{w}_{ij} + \mu_{ij2}) \nonumber\\
		&+ B^2_{ij}\mu_{i(j-1)2} + B^3_{ij}\mu_{i(j+1)2} \nonumber\\
	 \mu_{ij4} 	=& k_{ij4}(\hat{v}_{ij} - v_{ij}) \label{eq_observer_2}
	\end{align}
	where
	\begin{align}
		&D^1_{ij}(v_{ij}) = 		
		 \left\{ \begin{array}{l}
		 -\frac{b}{m_{ij}}v_{ij} - (c_{1}v_{ij} + c_{2}v_{ij}^2)  ,\  j = 1,M_i \\
		 -\frac{2b}{m_{ij}}v_{ij} - (c_{1}v_{ij} + c_{2}v_{ij}^2)  ,\  j = 2,\cdots,M_i-1 
		\end{array}\right. \nonumber\\
		&D^2_{ij}(v_{i(j-1)}) =  \left\{ \begin{array}{l}
		 0   ,\  j = 1 \\
		 \frac{b}{m_{ij}}v_{i(j-1)}   ,\  j = 2,\cdots,M_i 
		\end{array}\right.  \nonumber\\
		&D^3_{ij}(v_{i(j+1)}) =  \left\{ \begin{array}{l}
		 \frac{b}{m_{ij}}v_{i(j+1)}   ,\  j = 1,\cdots,M_i-1 \\
		 0   ,\  j = M_i
		\end{array}\right. .
	\end{align}
 
The main result is stated in the following theorem. 

	\begin{thm} \label{theo_observer}
		Consider the system composed of \eqref{eq_sys_1_observed} and the observer \eqref{eq_observer_1} with the auxiliary inputs designed in \eqref{eq_observer_2}. The parameters $k_{ij1} > 0$ and $K_{ij} = [k_{{ij}2},k_{{ij}3},k_{{ij}4}^T]^T$ are selected such that the matrix $(A_{ij} + K_{ij}C_{ij})$ is Hurwitz. Then, estimation is  asymptotically achieved in the sense of 
		\begin{align}
		\lim_{t \rightarrow \infty} [\chi_{ij}(t) -\hat\chi_{ij}(t)] = 0,\; (i,j)\in \C.
		\end{align}
	\end{thm}
	
	\begin{pf}
		Let $e_{p{ij}} = \hat{p}_{{ij}} - p_{ij}$, $p \in \{x,v,w,f \}$.
		The error dynamics can be obtained as follows,
		\begin{align*}
			\dot{e}_{x{ij}} =& e_{v{ij}} + \mu_{{ij}1} \nonumber\\
			\dot{e}_{v{ij}} =& e_{w{ij}} + \mu_{{ij}2} \nonumber\\
			\dot{e}_{w{ij}} =& B^1_{ij}(v_{ij})e_{w{ij}} + B^2_{ij}e_{w{i(j-1)}} \nonumber\\
			& + B^3_{ij}e_{w{i(j+1)}} + C_{ij}e_{f{ij}} + \mu_{{ij}3}\nonumber\\
			\dot{e}_{f{ij}} =& S_{ij}e_{f{ij}} + \mu_{{ij}4}
		\end{align*}
		Let $\zeta_{ij} = e_{w{ij}} + \mu_{{ij}2} - k_{{ij}2}e_{v{ij}}$.
		It is calculated that
		\begin{align*}
			\frac{d\mu_{{ij}2}(t)}{dt}  = & B^1_{ij}(v_{ij})\dot{v}_{{ij}} - B^1_{ij}(\hat{v}_{ij})\dot{\hat{v}}_{{ij}} \nonumber\\
			&+ B^2_{ij}\dot{v}_{{i(j-1)}} - B^2_{ij}\dot{\hat{v}}_{{i(j-1)}} \nonumber\\
			&+ B^3_{ij}\dot{v}_{{i(j+1)}} - B^3_{ij}\dot{\hat{v}}_{{i(j+1)}} \nonumber\\
			&+ k_{{ij}2}(\dot{\hat{v}}_{{ij}} - \dot{v}_{{ij}}).
		\end{align*}
		Then, 
		\begin{align}
			\dot{\zeta}_{ij} = & B^1_{ij2}(v_{ij})e_{w{ij}} + B^2_{ij}e_{w{i(j-1)}} \nonumber\\
			& + B^3_{ij}e_{w{i(j+1)}} + C_{ij}e_{f{ij}} + \mu_{{ij}3}  \nonumber\\
			&+ B^1_{ij2}(v_{ij})\dot{y}_{{ij}} - B^1_{ij2}(\hat{v}_{ij})\dot{\hat{y}}_{{ij}} \nonumber\\
			&+ B^2_{ij}\dot{v}_{{i(j-1)}} - B^2_{ij}\dot{\hat{v}}_{{i(j-1)}} \nonumber\\
			&+ B^3_{ij}\dot{v}_{{i(j+1)}} - B^3_{ij}\dot{\hat{v}}_{{i(j+1)}} \nonumber\\
			&+ k_{{ij}2}(\dot{\hat{v}}_{{ij}} - \dot{v}_{{ij}}) - k_{{ij}2}(e_{w{ij}} + \mu_{{ij}2} ) . \label{eq_dzeta_0}
		\end{align}
		Substituting the third equation of \eqref{eq_observer_2} into \eqref{eq_dzeta_0} gives
		\begin{align*}
			\dot{\zeta}_{ij} = k_{ij3}e_{v{ij}} + C_{ij}e_{f{ij}} 
		\end{align*}
		and
		\begin{align*}
			\dot{e}_{v{ij}} =& e_{w{ij}} + \mu_{{ij}2} \nonumber\\
			=& \zeta_{ij} - \mu_{{ij}2}+ k_{{ij}2}e_{v{ij}} + \mu_{{ij}2} \nonumber\\
			=& k_{{ij}2}e_{v{ij}} + \zeta_{ij}.
		\end{align*}
		From above, one has
		\begin{align*}
			\dot{e}_{v{ij}} =& k_{{ij}2}e_{v{ij}} + \zeta_{ij} \nonumber\\
			\dot{\zeta}_{ij} =& k_{{ij}3}e_{v{ij}} + C_{ij}e_{f{ij}} \nonumber\\
			\dot{e}_{f{ij}} =& k_{{ij}4}e_{v{ij}} + S_{ij}e_{f{ij}}.
		\end{align*}
		Define $\xi_{ij} = [e_{v{ij}},\zeta_{ij},e_{f{ij}}]^T$ that is governed by
		\begin{align}
			\dot{\xi}_{ij} = \begin{bmatrix} k_{{ij}2} & 1 & \textbf{0} \\ k_{{ij}3} & 0 & C_{ij} \\ k_{{ij}4} & \textbf{0} & S_{ij} \end{bmatrix} \xi_{ij}  = (A_{ij} + K_{ij}C_{ij})\xi_{ij} \label{dxi}
		\end{align}
		with
		\begin{align}
			A_{ij} = \begin{bmatrix} 0 & 1 & \textbf{0} \\ 0 & 0 & C_{ij} \\ \textbf{0} & \textbf{0} & S_{ij} \end{bmatrix},\ C_{ij} = \begin{bmatrix} 1 & 0 & \textbf{0} \end{bmatrix},\ K_{ij} = \begin{bmatrix} k_{{ij}2} \\ k_{{ij}3} \\ k_{{ij}4} \end{bmatrix},
		\end{align}
		where $\textbf{0}$ is the zero matrix of appropriate dimension. It can be directly verified that the pair $(A_{ij},C_{ij})$ is observable, therefore $(A_{ij} + K_{ij}C_{ij})$ can be configured as a Hurwitz matrix.
Together with 
		\begin{align*}
			\dot{e}_{x{ij}} =& e_{v{ij}} + \mu_{{ij}1}  = -k_{{ij}1}e_{x{ij}},
		\end{align*}
\eqref{dxi} becomes	\begin{align}
			\begin{bmatrix} \dot{e}_{x{ij}} \\ \dot{\xi}_{ij} \end{bmatrix} = D_{ij}\begin{bmatrix} e_{x{ij}} \\ \xi_{ij} \end{bmatrix},\; 
			D_{ij} = \begin{bmatrix} -k_{{ij}1} & \textbf{0} \\ \textbf{0} & (A_{ij} + K_{ij}C_{ij}) \end{bmatrix}.		 
			\label{eq_error_all}
		\end{align}
As the matrix $D_{ij}$ is Hurwitz, the system \eqref{eq_error_all} is stable and 
$\lim_{t \rightarrow \infty} e_{x{ij}}(t) = 0$ and $\lim_{t \rightarrow \infty} \xi_{ij}(t) = 0$.
The latter implies $\lim_{t \rightarrow \infty} e_{v{ij}}(t) = 0$, $\lim_{t \rightarrow \infty} e_{f{ij}}(t) = 0$, 
and $\lim_{t \rightarrow \infty} \zeta_{{ij}}(t) = 0$. From the definition of $\zeta$ and $\mu_{{ij}2}$, one has 
$\lim_{t \rightarrow \infty} \mu_{{ij}2}(t) = 0$ and 
 $\lim_{t \rightarrow \infty} e_{w{ij}}(t) = 0$.  
		The proof is thus completed.
	\end{pf}

	\begin{remark} \label{remark_mu}
		Due to the ingenious observer \eqref{eq_observer_1}, the estimation errors $e_{p{ij}}$, $p \in \{x,v,w,f \}$ asymptotically converge to zero regardless of the input $u_{ij}$.
		Therefore, the terms $\mu_{ijl},l =1,2,3,4$ in \eqref{eq_observer_1} satisfy $\lim_{t \rightarrow \infty} \mu_{ijl}(t) = 0$ for any input $u_{ij}$.
	\end{remark}

	\section{Controller Design} \label{sec_controller}
	
Based on the estimated states obtained by the observers in Section~\ref{sec_observer}, we aim 
to design a controller for every carriage to achieve the requirements ${\bf R1}$, ${\bf R2}$, and ${\bf R3}$ in this section. 
We first give the explicit construction of the controller and then the rigorous analysis in the proof of the main theorem. 
Two different types of controllers for the carriages in $\C_o$ and $\C_h$ are given below, separately. 

 For $(i,j)\in \C_o$, we first define the following three functions 
	\begin{align*}
& \alpha_{ij1}(\hat{x}_{ij}, \hat{x}_{i(j-1)}, \hat{v}_{i(j-1)})  \nonumber\\
& \alpha_{ij2}(\hat{x}_{ij}, \hat{x}_{i(j-1)}, \hat{v}_{ij}, \hat{v}_{i(j-1)}, \hat{w}_{i(j-1)})   \nonumber\\
 & \alpha_{ij3}(\hat{x}_{ij}, \hat{x}_{i(j-1)}, \hat{v}_{ij}, \hat{v}_{i(j-1)}, \hat{w}_{ij} ,\hat{w}_{i(j-1)}, \\
 &\;\;\;\;\;\; \;\;\;\; x_{ij},v_{ij},x_{i(j-1)},v_{i(j-1)},u_{i(j-1)}).
	\end{align*}
To keep the notation neat, we simply use $(\cdot)$ to represent the arguments
when the functions appear below. Specifically, the functions can be recursively defined  as
			\begin{align} 
 \alpha_{ij1}(\cdot) =& \hat{v}_{i(j-1)} - (l_{ij1} + 1)z_{ij1} \label{eq_z_alpha_c_2} \\
		 \alpha_{ij2}( \cdot ) = & -l_{ij2}z_{ij2}   - z_{ij1} - \frac{1}{2}z_{ij2}  \nonumber \\
		& + \frac{\partial \alpha_{ij1}(\cdot)}{\partial \hat{x}_{ij}}\hat{v}_{ij} - \frac{1}{2}(\frac{\partial \alpha_{ij1}(\cdot)}{\partial \hat{x}_{ij}})^2z_{ij2} \nonumber\\
		& + \frac{\partial \alpha_{ij1}(\cdot)}{\partial \hat{x}_{i(j-1)}}\hat{v}_{i(j-1)} - \frac{1}{2}(\frac{\partial \alpha_{ij1}(\cdot)}{\partial \hat{x}_{i(j-1)}})^2z_{ij2} \nonumber\\
		& + \frac{\partial \alpha_{ij1}(\cdot)}{\partial \hat{v}_{i(j-1)}}\hat{w}_{i(j-1)} - \frac{1}{2}(\frac{\partial \alpha_{ij1}(\cdot)}{\partial \hat{v}_{i(j-1)}})^2z_{ij2}   \label{eq_z_alpha_c_3} \\
 \alpha_{ij3}(\cdot)  =& -l_{ij3}z_{ij3} - z_{ij2}  + \frac{\partial \alpha_{ij2}(\cdot)}{\partial \hat{x}_{ij}} \dot{\hat{x}}_{ij} \nonumber \\
 & + \frac{\partial \alpha_{ij2}(\cdot)}{\partial \hat{x}_{i(j-1)}} \dot{\hat{x}}_{i(j-1)} + \frac{\partial \alpha_{ij2}(\cdot)}{\partial \hat{v}_{ij}} \dot{\hat{v}}_{ij} \nonumber\\
		& + \frac{\partial \alpha_{ij2}(\cdot)}{\partial \hat{v}_{i(j-1)}} \dot{\hat{v}}_{i(j-1)} + \frac{\partial \alpha_{ij2}(\cdot)}{\partial \hat{w}_{i(j-1)}} \dot{\hat{w}}_{i(j-1)}, \label{eq_z_alpha_c_4}
	\end{align}
together with	
	\begin{align} 
		z_{ij1} =& \hat{x}_{ij} - \hat{x}_{i(j-1)} + d_p \nonumber\\
		z_{ij2} =& \hat{v}_{ij} - \alpha_{ij1}(\cdot)  \nonumber\\
z_{ij3} =& \hat{w}_{ij} -\alpha_{ij2}(\cdot). \label{z}
	\end{align}
It is noted that the time derivatives in \eqref{eq_z_alpha_c_4} can be replaced by those in \eqref{eq_observer_1}.
Now, the controller is designed as 
			\begin{align}
		u_{ij} = & - [B^1_{ij}(v_{ij})\hat{w}_{ij} + B^2_{ij}\hat{w}_{i(j-1)} + B^3_{ij}\hat{w}_{i(j+1)} \nonumber\\
		& + C_{ij}\hat{f}_{ij} + \mu_{{ij}3}]   +\alpha_{ij3}( \cdot)    ,\;  (i,j) \in \C_o. \label{eq_u_p_ij_1}
	\end{align}

 For $(i,j)\in \C_h$, we define the following error transformation functions
		\begin{align}
		\phi_{i}(\tilde{x}_{i}) =& {\rm ln}\left( \frac{\rho_1\rho_2 + \rho_1\tilde{x}_{i}}{\rho_1\rho_2 - \rho_2\tilde{x}_{i}} \right) \nonumber\\
		\psi_{i}(\tilde{q}_{i}) =& {\rm ln}\left( \frac{\varrho_1\varrho_2 + \varrho_1\tilde{q}_{i}}{\varrho_1\varrho_2 - \varrho_2\tilde{q}_{i}} \right), \label{eq_two_error_trans}
		\end{align}
and hence
			\begin{align}
\Phi_i (\tilde{x}_{i} )=\frac{\partial \phi_{i} (\tilde{x}_{i})}{\partial \tilde{x}_{i}} ,\;
\Psi_i( \tilde{q}_{i} ) = \frac{\partial \psi_{i} (\tilde{q}_{i}) }{\partial \tilde{q}_{i}} . 
		\end{align}
Also, we define the functions
		\begin{align}
		\beta_{i1}(\tilde{x}_i,\tilde{v}_i) =& -\phi_{i} (\tilde{x}_{i} )\Phi_i (\tilde{x}_{i} ) - \ell_{i2}\tilde{q}_{i} - \ell_{i3}\psi_{i}(\tilde{q}_{i}) \Psi_i (\tilde{q}_{i})\nonumber\\
		\beta_{i2} (\tilde{x}_i,\tilde{v}_i,\tilde{\hat{w}}_{i} )  =& \tilde{\hat{w}}_{i} + \ell_{i1}\tilde{v}_{i} -\beta_{i1}(\tilde{x}_i,\tilde{v}_i)  \label{eq_z_alpha_e_1}
		\end{align}
where $\tilde{\hat{w}}_{i} = \hat{w}_{(i-1)M_{i-1}} - \hat{w}_{i1}$ and, for $i=1$, 
$\hat{w}_{(i-1)M_{i-1}} = w_0$. Also, we simply use $(\cdot)$ to represent the arguments in these two functions when they appear below.  Now, the controller is designed as 
		\begin{align}
			u_{i1} =& G_{i-1} - [B^1_{i1}(v_{i1})\hat{w}_{i1} + B^3_{i1}\hat{w}_{i2} + C_{i1} \hat{f}_{i1} + \mu_{{i1}3} ] \nonumber\\
			& + \ell_{i1}\tilde{\hat{w}}_{i} + \ell_{i1}^2\beta_{i2}(\cdot) - \frac{\partial \beta_{i1}(\cdot)}{\partial \tilde{x}_{i}}\tilde{v}_{i} - \frac{\partial \beta_{i1}(\cdot)}{\partial \tilde{v}_{i}}\tilde{\hat{w}}_{i} \nonumber\\
			& + \left[\frac{\partial \beta_{i1}(\cdot)}{\partial \tilde{v}_{i}}\right]^2\beta_{i2}(\cdot) + \tilde{q}_{i} - \ell_{i4}\beta_{i2}(\cdot), \; i \in \T	 \label{eq_u_p_ij_3}
		\end{align}
		where
		\begin{align*}
		G_{i-1} =  \left\{ \begin{array}{ll}
		 u_0  , &  i = 1 \\
		 g_{i-1}  ,&  i = 2,\cdots,N 
		 \end{array}\right. ,
		\end{align*}
		and
		\begin{align}
		g_{i-1} =& B^1_{(i-1)M_{i-1}}(v_{(i-1)M_{i-1}})\hat{w}_{(i-1)M_{i-1}} \nonumber\\
		& + B^2_{(i-1)M_{i-1}}\hat{w}_{(i-1)(M_{i-1}-1)} + C_{(i-1)M_{i-1}}\hat{f}_{(i-1)M_{i-1}} \nonumber\\
		& + u_{(i-1)M_{i-1}} + \mu_{{(i-1)M_{i-1}}3}. \nonumber
		\end{align}

%
%

The main result is stated in the following theorem.

	\begin{thm} \label{theorem_con}
Consider the system~\eqref{eq_sys_1_observed}, the observer~\eqref{eq_observer_1}
given in Theorem~\ref{theo_observer}, and the controllers 
\eqref{eq_u_p_ij_1} and \eqref{eq_u_p_ij_3}.  Suppose the control parameters satisfy
\begin{align}
& l_{ij1} > 0,\; l_{ij2} > 0, \; l_{ij3} > 0 \nonumber\\
&0<\ell_{i1} < \min\{\sigma_2/\rho_1, \sigma_1/\rho_2\}\nonumber\\
&\ell_{i2} > 2,\;  \ell_{i3}> 2 + \ell_{i2}^2/2 ,\; \ell_{i4} > 1/2 \label{parameter}
\end{align}
and the initial conditions satisfy
 \begin{align}
 -\rho_2 < \tilde{x}_{i}(0) < \rho_1,\; 
 -\varrho_{2} < \tilde{q}_{i}(0) < \varrho_{1},\; i \in \T. \label{initial}
\end{align}
Then, the closed-loop system achieves the  requirements ${\bf R1}$, ${\bf R2}$, and ${\bf R3}$. 			
	\end{thm}

 \begin{pf} We first consider the carriages $(i,j)\in \C_o$.
	 	Define  a positive definite function as follows:
	\begin{align}
		V_{c1} = \frac{1}{2}\sum_{i = 1}^{N}\sum_{j = 2}^{M_i}z_{ij1}^2
	\end{align}
	then
	\begin{align}
		\dot{V}_{c1} =& \sum_{i = 1}^{N}\sum_{j = 2}^{M_i}z_{ij1}(\hat{v}_{ij} - \hat{v}_{i(j-1)} + \mu_{ij1} - \mu_{i(j-1)1}) \nonumber\\
		\leq& \sum_{i = 1}^{N}\sum_{j = 2}^{M_i}z_{ij1}(\hat{v}_{ij} - \hat{v}_{i(j-1)} + z_{ij1}) \nonumber\\
		& + \frac{1}{2}\sum_{i = 1}^{N}\sum_{j = 2}^{M_i}( \mu_{ij1}^2 + \mu_{i(j-1)1}^2).
	\end{align} 
	Define another positive definite function as follows:
	\begin{align}
	V_{c2} = V_{c1} + \frac{1}{2}\sum_{i = 1}^{N}\sum_{j = 2}^{M_i}z_{ij2}^2
	\end{align}
	then
	\begin{align}
		\dot{V}_{c2} \leq& \sum_{i = 1}^{N}\sum_{j = 2}^{M_i}-l_{ij1}z_{ij1}^2 + \frac{1}{2}\sum_{i = 1}^{N}\sum_{j = 2}^{M_i}( \mu_{ij1}^2 + \mu_{i(j-1)1}^2) \nonumber\\
		& + \sum_{i = 1}^{N}\sum_{j = 2}^{M_i}z_{ij2}[z_{ij1} + \hat{w}_{ij} + \mu_{ij2} - \frac{\partial \alpha_{ij1}(\cdot)}{\partial \hat{x}_{ij}}(\hat{v}_{ij} + \mu_{ij1}) \nonumber\\
		& - \frac{\partial \alpha_{ij1}(\cdot)}{\partial \hat{x}_{i(j-1)}}(\hat{v}_{i(j-1)} + \mu_{i(j-1)1}) \nonumber\\
		& - \frac{\partial \alpha_{ij1}(\cdot)}{\partial \hat{v}_{i(j-1)}}(\hat{w}_{i(j-1)} + \mu_{i(j-1)2})] \nonumber\\
		\leq& \sum_{i = 1}^{N}\sum_{j = 2}^{M_i}-l_{ij1}z_{ij1}^2 + \frac{1}{2}\sum_{i = 1}^{N}\sum_{j = 2}^{M_i}( \mu_{ij1}^2 + \mu_{i(j-1)1}^2) \nonumber\\
		& + \sum_{i = 1}^{N}\sum_{j = 2}^{M_i}z_{ij2}[\hat{w}_{ij} + z_{ij1} + \frac{1}{2}z_{ij2} \nonumber\\
		& - \frac{\partial \alpha_{ij1}(\cdot)}{\partial \hat{x}_{ij}}\hat{v}_{ij} + \frac{1}{2}(\frac{\partial \alpha_{ij1}(\cdot)}{\partial \hat{x}_{ij}})^2z_{ij2} \nonumber\\
		& - \frac{\partial \alpha_{ij1}(\cdot)}{\partial \hat{x}_{i(j-1)}}\hat{v}_{i(j-1)} + \frac{1}{2}(\frac{\partial \alpha_{ij1}(\cdot)}{\partial \hat{x}_{i(j-1)}})^2z_{ij2} \nonumber\\
		& - \frac{\partial \alpha_{ij1}(\cdot)}{\partial \hat{v}_{i(j-1)}}\hat{w}_{i(j-1)} + \frac{1}{2}(\frac{\partial \alpha_{ij1}(\cdot)}{\partial \hat{v}_{i(j-1)}})^2z_{ij2} ] \nonumber\\
		& + \frac{1}{2}\sum_{i = 1}^{N}\sum_{j = 2}^{M_i}[\mu_{ij2}^2 + \mu_{ij1}^2 + \mu_{i(j-1)1}^2 + \mu_{i(j-1)2}^2].
	\end{align}
Define the Lyapunov candidate function as follows:
	\begin{align}
	V_{c3} = V_{c2} + \frac{1}{2}\sum_{i = 1}^{N}\sum_{j = 2}^{M_i}z_{ij3}^2
	\end{align}
	then
	\begin{align}
		\dot{V}_{c3} \leq& \sum_{i = 1}^{N}\sum_{j = 2}^{M_i}[-l_{ij1}z_{ij1}^2 - l_{ij2}z_{ij2}^2] \nonumber\\
		& + \sum_{i = 1}^{N}\sum_{j = 2}^{M_i}z_{ij3}[z_{ij2} +  \dot{\hat{w}}_{ij} - \dot{\alpha}_{ij2}(\cdot)] \nonumber\\
		& + \frac{1}{2}\sum_{i = 1}^{N}\sum_{j = 2}^{M_i}[\mu_{ij2}^2 + 2\mu_{ij1}^2 + 2\mu_{i(j-1)1}^2 + \mu_{i(j-1)2}^2] .\label{eq_d_Vc_3_1}
	\end{align}

%
 
 Substituting \eqref{eq_z_alpha_c_4} and \eqref{eq_u_p_ij_1} into \eqref{eq_d_Vc_3_1} gives
	\begin{align}
		\dot{V}_{c3} \leq& \sum_{i = 1}^{N}\sum_{j = 2}^{M_i}[-l_{ij1}z_{ij1}^2 - l_{ij2}z_{ij2}^2 - l_{ij3}z_{ij3}^2] \nonumber\\
	 & + \frac{1}{2}\sum_{i = 1}^{N}\sum_{j = 2}^{M_i}[\mu_{ij2}^2 + 2\mu_{ij1}^2 + 2\mu_{i(j-1)1}^2 + \mu_{i(j-1)2}^2] \nonumber \\
		 \leq& -2p_cV_{c3} + \Omega_c \label{eq_d_Vc_3_3}
	\end{align}
for
	\begin{align}
	p_c =& \min_{(i,j)\in \C_o} \left\{ l_{ij1}, l_{ij2}, l_{ij3} \right\} \nonumber\\
	\Omega_c =& \frac{1}{2}\sum_{i = 1}^{N}\sum_{j = 2}^{M_i}[\mu_{ij2}^2 + 2\mu_{ij1}^2 + 2\mu_{i(j-1)1}^2 + \mu_{i(j-1)2}^2].
	\end{align}
	
As $\lim_{t \rightarrow \infty} \mu_{ijl} (t)= 0,\; l = 1,2,3,4$ holds by Remark~\ref{remark_mu}, so does $\lim_{t \rightarrow \infty}\Omega_c(t) = 0$. 	For any constant $\varsigma>0$, there exists a finite time $T_1>0$ such that $\Omega_c (t) \leq \varsigma p_c$, $\forall t\geq T_1$. It implies that, for $t \geq T_1$,
	\begin{align}
	\dot{V}_{c3} \leq -2p_cV_{c3} + \varsigma p_c \label{eq_d_Vc_3_4}
	\end{align}
	By the comparison principle, for $t\geq T_1$,
	\begin{align}
	V_{c3}(t) \leq V_{c3}(T_1) e^{-2p_c(t - T_1)} + \frac{\varsigma}{2}\left(1 - e^{-2p_c(t - T_1)}\right)
	\end{align}
	Therefore, there exists a finite time $T_2 > T_1$ such that $V_{c3}(t) < \varsigma$, $\forall t\geq T_2$. It concludes $\lim_{t \rightarrow \infty}V_{c3}(t) = 0$ and hence $\lim_{t \rightarrow \infty}z_{ij1}(t) = 0$, $\lim_{t \rightarrow \infty}z_{ij2}(t) = 0$ and $\lim_{t \rightarrow \infty}z_{ij3}(t) = 0$. Thus, $\lim_{t\rightarrow \infty} [\hat{x}_{i(j-1)}(t) - \hat{x}_{ij}(t)] = d_p$ by \eqref{z},
$\lim_{t\rightarrow \infty}[\alpha_{ij1}(\cdot (t)) - \hat{v}_{i(j-1)}(t)]=0$ by \eqref{eq_z_alpha_c_2}, and  
$\lim_{t\rightarrow \infty}[\alpha_{ij1}(\cdot (t)) - \hat{v}_{ij}(t)]=0$  by \eqref{z}. The last two equations also give $\lim_{t\rightarrow \infty} [\hat{v}_{i(j-1)}(t) - \hat{v}_{ij}(t)] = 0$. By Theorem~\ref{theo_observer}, one has $\lim_{t\rightarrow \infty} [x_{ij}(t) - \hat{x}_{ij}(t)] = 0$ and $\lim_{t\rightarrow \infty} [v_{ij}(t) - \hat{v}_{ij}(t)] = 0$. As a result,  $\lim_{t\rightarrow \infty} [x_{i(j-1)}(t) - x_{ij}(t)] = d_p$ and $\lim_{t\rightarrow \infty} [v_{i(j-1)}(t) - v_{ij}(t)] = 0$,  $(i,j)\in \C_o$, that is,   {\bf R1} is proved.

\medskip

Next, we consider the carriages $(i,j)\in \C_h$. The error transformation function \eqref{eq_two_error_trans} has the following properties.
		
		\begin{itemize}
			\item[(P1)] If $-\rho_2 < \tilde{x}_i(0) < \rho_1$ and $\phi_{i}(\tilde{x}_{i}(t))$ is bounded for $t > 0$, then $-\rho_2 < \tilde{x}_i(t) < \rho_1$ holds for $t>0$.  The same property holds for $\psi_{i}(\tilde{q}_{i}(t))$ and $\tilde{q}_{i}(t)$.
			\item[(P2)] (Refer to \cite{cao2020performance})  there exist positive constants $\theta_{i}$, $\vartheta_{i}$ such that the following inequalities hold
			\begin{align}
			\tilde{x}_{i}\phi_{i}(\tilde{x}_{i})\Phi_i(\tilde{x}_{i}) &\geq \theta_{i}\phi_{i}^2 (\tilde{x}_{i}) \nonumber\\
			\tilde{q}_{i}\psi_{i}(\tilde{q}_{i})  \Psi_i (\tilde{q}_{i})  &\geq \vartheta_{i}\psi_{i}^2(\tilde{q}_{i}) . \label{eq_proper_2}
			\end{align}
			\item[(P3)] $\Phi_i(\tilde{x}_i)>0$ for $-\rho_2 < \tilde{x}_i < \rho_1$; and $\lim_{t \rightarrow 0}\phi_{i}(\tilde{x}_{i}(t)) = 0$ implies $\lim_{t \rightarrow 0}\tilde{x}_{i}(t) = 0$. The same properties hold for $\Psi_i(\tilde{q}_{i})$, $\psi_{i}(\tilde{q}_{i})$, and $\tilde{q}_{i}$.
		\end{itemize}

 The remaining proof is divided into two parts. In part (a), we will prove  $\tilde{x}_{i}(t) \in (-\rho_2, \rho_1)$, $\forall t > 0$, $\lim_{t \rightarrow \infty}\tilde{x}_{i}(t) = 0$, and $\lim_{t \rightarrow \infty}\tilde{v}_{i}(t) = 0$; and in part (b), we will prove that $\tilde{v}_{i}(t)$ is within the prescribed range $(-\sigma_2, \sigma_1)$, $\forall t>0$.
		
		(a) Consider a Lyapunov candidate function
		\begin{align}
		V_{e1} =& \frac{1}{2}\sum_{i = 2}^{N}\tilde{q}_{i}^2 + \frac{1}{2}\sum_{i = 2}^{N}\phi_{i}^2(\tilde{x}_{i}) + \frac{1}{2}\sum_{i = 2}^{N}\beta^2_{i2}(\cdot) \label{eq_Ve_1},
		\end{align}
		whose time derivative satisfies
		\begin{align}
		\dot{V}_{e1} =& \sum_{i = 2}^{N}\tilde{q}_{i}\dot{\tilde{q}}_{i} + \sum_{i = 2}^{N}\phi_{i}(\tilde{x}_{i}) \dot{\phi}_{i} (\tilde{x}_{i}) + \sum_{i = 2}^{N}\beta_{i2}(\cdot)\dot{\beta}_{i2}(\cdot) \nonumber\\
		=& \sum_{i = 2}^{N}\tilde{q}_{i}\left[\tilde{w}_{i} + \ell_{i1}\tilde{v}_{i}\right] + \sum_{i = 2}^{N}\phi_{i}(\tilde{x}_{i})  \Phi_i(\tilde{x}_{i})  \tilde{v}_{i} \nonumber\\
		&+ \sum_{i = 2}^{N}\beta_{i2}(\cdot)[\dot{\tilde{\hat{w}}}_{i} + \ell_{i1}\dot{\tilde{v}}_{i} - \dot{\beta}_{i1}(\cdot)] \nonumber\\
		=& \sum_{i = 2}^{N}\tilde{q}_{i}\left[\beta_{i2}(\cdot) + \beta_{i1}(\cdot) - e_{w(i-1)M_{i-1}} + e_{wi1}\right] \nonumber\\
		&+ \sum_{i = 2}^{N}\phi_{i}(\tilde{x}_{i}) \Phi_i(\tilde{x}_{i})  \left[\tilde{q}_{i} - \ell_{i1}\tilde{x}_{i} \right] \nonumber\\
		&+ \sum_{i = 2}^{N}\beta_{i2}(\cdot)[\dot{\hat{w}}_{(i-1)M_{i-1}} - \dot{\hat{w}}_{i1} + \ell_{i1}\dot{\tilde{v}}_{i} - \dot{\beta}_{i1}(\cdot)] \nonumber\\
		\leq& \sum_{i = 2}^{N}\tilde{q}_{i}\left[ \beta_{i1}(\cdot) + \phi_{i}(\tilde{x}_{i}) \Phi_i (\tilde{x}_{i}) + \tilde{q}_{i}\right] \nonumber\\
		&+ \sum_{i = 2}^{N}\left[- \ell_{i1}\tilde{x}_{i}\phi_{i}(\tilde{x}_{i}) \Phi_i (\tilde{x}_{i}) \right] \nonumber\\& + \frac{1}{2}\sum_{i = 2}^{N}\left[ e_{w(i-1)M_{i-1}}^2 + e_{wi1}^2 \right] \nonumber\\
		&+ \sum_{i = 2}^{N}\beta_{i2}(\cdot)[\dot{\hat{w}}_{(i-1)M_{i-1}} - \dot{\hat{w}}_{i1} + \ell_{i1}\tilde{\hat{w}}_{i} \nonumber\\
		& + \ell_{i1}( - e_{w(i-1)M_{i-1}} + e_{wi1}) - \frac{\partial \beta_{i1}(\cdot)}{\partial \tilde{x}_{i}}\tilde{v}_{i} - \frac{\partial \beta_{i1}(\cdot)}{\partial \tilde{v}_{i}}\tilde{\hat{w}}_{i} \nonumber\\
		& - \frac{\partial \beta_{i1}(\cdot)}{\partial \tilde{v}_{i}}( - e_{w(i-1)M_{i-1}} + e_{wi1}) + \tilde{q}_{i} ]  \label{eq_d_Ve_1_1}
		\end{align}

According to \eqref{eq_observer_1}, one has
		\begin{align}
		\dot{\hat{w}}_{(i-1)M_{i-1}} - \dot{\hat{w}}_{i1} = G_{i-1} - [B^1_{i1}(v_{i1})\hat{w}_{i1} \nonumber\\ 
		  + B^3_{i1}\hat{w}_{i2} + C_{i1} \hat{f}_{i1} + u_{i1} + \mu_{{i1}3}] . \label{eq_dw_e_1}
		\end{align}
Using \eqref{eq_dw_e_1} and the following facts
		\begin{align*}
			&\sum_{i = 2}^{N}  \beta_{i2}(\cdot)\ell_{i1}( - e_{w(i-1)M_{i-1}} + e_{wi1}) \nonumber\\
			  \leq & \sum_{i = 2}^{N}  \beta_{i2}^2(\cdot)\ell_{i1}^2 + \frac{1}{2}\sum_{i = 2}^{N}(e_{w(i-1)M_{i-1}}^2 + e_{wi1}^2) \nonumber\\
			&\sum_{i = 2}^{N}  -\beta_{i2}(\cdot) \frac{\partial \beta_{i1}(\cdot)}{\partial \tilde{v}_{i}}( - e_{w(i-1)M_{i-1}} + e_{wi1}) \nonumber\\
			  \leq& \sum_{i = 2}^{N}  \beta_{i2}^2(\cdot)\left[\frac{\partial \beta_{i1}(\cdot)}{\partial \tilde{v}_{i}}\right]^2 + \frac{1}{2}\sum_{i = 2}^{N}(e_{w(i-1)M_{i-1}}^2 + e_{wi1}^2), 
		\end{align*}
  together with the definitions of $\beta_{i1}(\cdot)$ and $u_{i1}$ in \eqref{eq_z_alpha_e_1} and \eqref{eq_u_p_ij_3}, and the property (P2), one has
%
%
%
		\begin{align}
		\dot{V}_{e1} \leq& \sum_{i = 2}^{N}\left[  - (\ell_{i2} - 1)\tilde{q}_{i}^2 - \ell_{i3}\vartheta_{i}\psi_{i}^2 (\tilde{q}_{i}) \right] + \sum_{i = 2}^{N}\left[- \ell_{i1}\theta_{i}\phi_{i}^2(\tilde{x}_{i}) \right] \nonumber\\
		&+ \frac{3}{2}\sum_{i = 2}^{N}\left[ e_{w(i-1)M_{i-1}}^2 + e_{wi1}^2 \right] - \ell_{i4}\beta^2_{i2}(\cdot) \nonumber\\
		 \leq& -2p_{e1}V_{e1} + \Omega_{e1} \label{eq_d_Ve_1_3}
		\end{align}		
for, noting \eqref{parameter},
		\begin{align*}
		p_{e1} =& \min_{i\in\T} \{(\ell_{i2} - 1), \ell_{i1}\theta_{i}, \ell_{i4} \} >0 \nonumber\\
		\Omega_{e1} =& \frac{3}{2}\sum_{i = 2}^{N}\left[ e_{w(i-1)M_{i-1}}^2 + e_{wi1}^2 \right].
		\end{align*}
		
		Integrating \eqref{eq_d_Ve_1_3} over $[0, t]$ yields
		\begin{align}
		0 \leq V_{e1}(t) \leq \frac{\Omega_{e1}}{2p_{e1}} + V_{e1}(0)e^{-2p_{e1}t}. \label{eq_Ve_1_1}
		\end{align}
It is known that $V_{e1}(0)$ is bounded for the initial values satisfying  \eqref{initial}.	
		It follows from \eqref{eq_Ve_1_1} that $\phi_{i}(\tilde{x}_{i}(t))$ is bounded and hence $\tilde{x}_{i}(t) \in (-\rho_2, \rho_1)$, 
		for $t > 0$,																				 due to the property (P1).

By Theorem~\ref{theo_observer}, one has $\lim_{t \rightarrow \infty} e_{w(i-1)M_{i-1}}(t) = 0$ and $\lim_{t \rightarrow \infty}e_{wi1} (t)= 0 $,  thus $\lim_{t \rightarrow \infty}\Omega_{e1}(t) = 0$. Therefore, the analysis on \eqref{eq_d_Ve_1_3} can be similar to that on \eqref{eq_d_Vc_3_3}, from which it is concluded that  $\lim_{t \rightarrow \infty}V_{e1}(t) = 0$. Furthermore, for the property (P3) with
 $\lim_{t \rightarrow \infty}\phi_{i}(\tilde{x}_{i}(t)) = 0$ and $\lim_{t \rightarrow \infty}\psi_{i}(\tilde{q}_{i}(t)) = 0$, one has $\lim_{t \rightarrow \infty}\tilde{x}_{i}(t) = 0$, $\lim_{t \rightarrow \infty}\tilde{q}_{i} (t)= 0$, and hence $\lim_{t \rightarrow \infty}\tilde{v}_{i}(t)= 0$.
		
		(b) Consider a Lyapunov candidate function
		\begin{align}
		V_{e2} =& \frac{1}{2}\sum_{i = 2}^{N}\tilde{q}_{i}^2 + \frac{1}{2}\sum_{i = 2}^{N}\psi_{i}^2(\tilde{q}_{i})+ \frac{1}{2}\sum_{i = 2}^{N}\beta_{i2}^2(\cdot) \label{eq_Ve_2}
		\end{align}
whose time derivative satisfies
		\begin{align}
		\dot{V}_{e2} =& \sum_{i = 2}^{N}\tilde{q}_{i}\dot{\tilde{q}}_{i} + \sum_{i = 2}^{N}\psi_{i}(\tilde{q}_{i})\dot{\psi}_{i} (\tilde{q}_{i})+ \sum_{i = 2}^{N}\beta_{i2}(\cdot)\dot{\beta}_{i2}(\cdot) \nonumber\\
		=& \sum_{i = 2}^{N} [\tilde{q}_{i} + \psi_{i}(\tilde{q}_{i})\Psi_i (\tilde{q}_{i})] \nonumber\\
		&\times \left[\beta_{i2}(\cdot) + \beta_{i1}(\cdot) - e_{w(i-1)M_{i-1}} + e_{wi1}\right] \nonumber\\
		&+ \sum_{i = 2}^{N}\beta_{i2}(\cdot)(\dot{\hat{w}}_{(i-1)M_{i-1}} - \dot{\hat{w}}_{i1} + \ell_{i1}\dot{\tilde{v}}_{i} - \dot{\beta}_{i1}(\cdot)) \nonumber\\
		\leq& \sum_{i = 2}^{N}\tilde{q}_{i}\left[ \beta_{i1}(\cdot) + \tilde{q}_{i} \right] + \frac{1}{2}\sum_{i = 2}^{N}[e_{w(i-1)M_{i-1}}^2 + e_{wi1}^2] \nonumber\\
		&+ \sum_{i = 2}^{N}\psi_{i}(\tilde{q}_{i})\Psi_i(\tilde{q}_{i})\left[ \beta_{i1}(\cdot) + \frac{3}{2}\psi_{i}(\tilde{q}_{i})\Psi_i(\tilde{q}_{i}) \right]  \nonumber \\
		&+ \frac{3}{2}\sum_{i = 2}^{N}[e_{w(i-1)M_{i-1}}^2   + e_{wi1}^2]  \nonumber\\
		&+ \sum_{i = 2}^{N}\beta_{i2}(\cdot) \Big[\dot{\hat{w}}_{(i-1)M_{i-1}} - \dot{\hat{w}}_{i1} + \ell_{i1}\tilde{\hat{w}}_{i} \nonumber\\
		& + \ell_{i1}^2\beta_{i2}(\cdot) - \frac{\partial \beta_{i1}(\cdot)}{\partial \tilde{x}_{i}}\tilde{v}_{i} - \frac{\partial \beta_{i1}(\cdot)}{\partial \tilde{v}_{i}}\tilde{\hat{w}}_{i} \nonumber\\
		& + \left[\frac{\partial \beta_{i1}(\cdot)}{\partial \tilde{v}_{i}}\right]^2\beta_{i2}(\cdot) + \tilde{q}_{i} + \frac{1}{2}\beta_{i2}(\cdot) \Big]. \label{eq_d_Ve_2_1}
		\end{align}
		Using \eqref{eq_dw_e_1} and  the definitions of $\beta_{i1}(\cdot)$ and $u_{i1}$ in \eqref{eq_z_alpha_e_1} and \eqref{eq_u_p_ij_3} gives 
		\begin{align}
		\dot{V}_{e2} \leq& \sum_{i = 2}^{N} \left[ - (\ell_{i2} - 2)\tilde{q}_{i}^2 - (\ell_{i3} - 2 - \frac{\ell_{i2}^2}{2})\left( \psi_{i}(\tilde{q}_{i})\Psi_i (\tilde{q}_{i})\right)^2 \right] \nonumber\\
		&+ \sum_{i = 2}^{N}\left[ -\ell_{i3}\vartheta_{i}\psi_{i}^2 (\tilde{q}_{i}) - (\ell_{i4} - \frac{1}{2})\beta_{i2}^2(\cdot) \right]\nonumber\\
		& + \sum_{i = 2}^{N}\left[ 2e_{w(i-1)M_{i-1}}^2 + 2e_{wi1}^2 +\left( \phi_{i}(\tilde{x}_{i}) \Phi_i (\tilde{x}_{i}) \right)^2 \right]\nonumber\\
		\leq& -2p_{e2}V_{e2} + \Omega_{e2} \label{eq_d_Ve_2_4} 
		\end{align}
for, noting  \eqref{parameter}, 	 
		\begin{align*}
		p_{e2} =& \min_{i\in\T} \left\{(\ell_{i2} - 2), \ell_{i3}\vartheta_{i}, (\ell_{i4} - \frac{1}{2})  \right\} >0 \nonumber\\
		\Omega_{e2} =& 2\sum_{i = 2}^{N}\left[ e_{w(i-1)}^2 + e_{wi}^2 + \left( \phi_{i}(\tilde{x}_{i}) \Phi_i (\tilde{x}_{i}) \right)^2 \right].
		\end{align*}
 
		Since $\lim_{t \rightarrow \infty}\tilde{x}_{i} (t)= 0$ and $\tilde{x}_{i} \in (-\rho_2, \rho_1)$   from part (a), and 
		$\phi_{i}(\tilde{x}_{i})\Phi_i(\tilde{x}_{i})$ is  continues in  $\tilde{x}_{i} \in (-\rho_2, \rho_1)$, one has 
		$\lim_{t \rightarrow \infty}\left( \phi_{i}(\tilde{x}_{i}(t))\Phi_i(\tilde{x}_{i}(t))\right)^2 = 0$ and 
		$\left( \phi_{i}(\tilde{x}_{i})\Phi_i(\tilde{x}_{i})\right)^2$ is bounded.
		It, together with  $\lim_{t \rightarrow \infty}(e_{w(i-1)}^2(t)+ e_{wi}^2(t)) = 0$, implies $\lim_{t \rightarrow \infty}\Omega_{e2} (t)= 0$.  	Therefore, the analysis on \eqref{eq_d_Ve_2_4} can be similar to that on \eqref{eq_d_Vc_3_3}, from which it is concluded that
		 $\lim_{t \rightarrow \infty}V_{e2}(t) = 0$. 
		 Then $\psi_{i}(\tilde{q}_{i})$ is bounded and hence $\tilde{q}_{i} \in (-\varrho_2, \varrho_1)$, $\forall t>0$ for the property (P1) of $\psi_{i}(\tilde{q}_{i})$. According to Remark~\ref{remark_definition_q}, 
		 $\tilde{q}_{i} \in (-\varrho_2, \varrho_1)$, together with $\tilde{x}_{i}(t) \in (-\rho_2, \rho_1)$ in part (a), gives $\tilde{v}_{i}(t) \in (-\sigma_2, \sigma_1)$, $\forall t > 0$ is obtained.
		
In part (a), one has proved $\lim_{t \rightarrow \infty}\tilde{x}_{i} (t)= 0$, 
$\lim_{t \rightarrow \infty}\tilde{v}_{i} (t)= 0$ and $\tilde{x}_{i}(t) \in (-\rho_2, \rho_1)$, $\forall t > 0$;
in part (b),   $\tilde{v}_{i}(t) \in (-\sigma_2, \sigma_1)$, $\forall t > 0$. The proof of the control requirements {\bf R2} and {\bf R3} 
is thus completed. \end{pf}
	
 \begin{remark} \label{remark_communi} 
In the  railway network system studied in this paper, we consider
two types of information exchange to facilitate implementation of the proposed design. 
In order to improve the flexibility of carriage assembly, 
it is assumed that carriages of a same train can only obtain the information of neighboring carriages 
through a train bus. For carriages of different trains, only adjacent trains can transmit information via wireless communication. 
That is, for two adjacent trains, the head carriage of the rear train can obtain the information from the tail carriage of the front train.
In Remark~\ref{remark_measurable}, $x_{ij}$ and $v_{ij}$ are measurable and  available for implementation of \eqref{varpiij}. 		
All the carriages can obtain the information of adjacent carriages, so the information $\hat{x}_{i(j-1)}, \hat{v}_{i(j-1)}, \hat{w}_{i(j-1)},x_{i(j-1)}, v_{i(j-1)}, u_{i(j-1)}$ of the $(j-1)$-th carriage can be obtained by the $j$-th carriage in $i$-th train for implementation of the observer \eqref{eq_observer_1} and the controllers \eqref{eq_u_p_ij_1} and \eqref{eq_u_p_ij_3}.

\end{remark}

	\section{Numerical Simulation} \label{sec_simulation}
%

	\begin{table}[t]
		\begin{center}
		\caption{Values of the train parameters}\label{tab_parameter}
				\begin{tabular}{lll}
				\hline
				\textbf{Parameter} & \textbf{Value} & \textbf{Unit} \\
				\hline
				$m_{ij}$ & $80 \times 10^3$ & kg \\
				$c_0$ & $0.01176$ & N/kg \\
				$c_1$ & $0.00077616$ & Ns/mkg \\
				$c_2$ & $1.6 \times 10^{-5}$ & ${\rm Ns^2/m^2kg}$ \\
				$a$ & $1.6 \times 10^{5}$ & N/m \\
				$b$ & $600$ & Ns/m \\
				$\upsilon_{ij}$ & $2 \times 10^5$ & / \\
				$\omega_{ij}$ & $1$ & rad/s \\
				$\nu_{ij}$ & $2 \times 10^5$ & s/rad \\
				$F_{cij}$ & $1$ & ${\rm N/s}$ \\
				$F_{pij}$ & $1$ & ${\rm N/s}$ \\
				$F_{\phi ij}$ & $ 6(i-1)+2j$ & rad \\
				$r_{ij}$ & $50$ & ${\rm s^{-1}}$ \\
				\hline
			\end{tabular}
		\end{center}
	\end{table}
	\begin{table*}[h]
		\begin{center}
			\caption{Initial values of the system states}
			\begin{tabular}{llllllllll}
				\hline
				$(i, j)$ & $(1, 1)$ & $(1, 2)$ & $(1, 3)$ & $(2, 1)$ & $(2, 2)$ & $(2, 3)$ & $(3, 1)$ & $(3, 2)$ & $(3, 3)$ \\
				\hline
				$x_{ij}(0)({\rm m})$ & $13062$ & $13036$ & $13010$ & $5157$ & $5131$ & $5105$ & $52$ & $26$ & $0$ \\
				$v_{ij}(0)({\rm m/s})$ & $20.5$ & $20.2$ & $20.3$ & $19.8$ & $19.9$ & $20.5$ & $19.7$ & $20.5$ & $20.2$ \\
				$w_{ij}(0)({\rm m/s^2})$ & $0$ & $0$ & $0$ & $0$ & $0$ & $0$ & $0$ & $0$ & $0$ \\
				\hline
			\end{tabular}\label{tab_state}
		\end{center}
	\end{table*}

	\begin{table*}[h]
		\begin{center}
			\caption{Occurrence intervals of actuator faults}
			\begin{tabular}{lllllllllll}
				\hline
				&$(i, j)$ & $(1, 1)$ & $(1, 2)$ & $(1, 3)$ & $(2, 1)$ & $(2, 2)$ & $(2, 3)$ & $(3, 1)$ & $(3, 2)$ & $(3, 3)$ \\
				\cline{2-11}
				{$f_{ij1}$} &$[t^s_{ij1},t^e_{ij1}] ({\rm \times 10^2\ s})$ & $[4, 14]$ & $[6, 15]$ & $[8, 16]$ & $[10, 17]$ & $[12, 18]$ & $[14, 19]$ & $[16, 20]$ & $[18, 21]$ & $[20, 22]$ \\
				{$f_{ij3}$}&$[t^s_{ij3},t^e_{ij3}] ({\rm \times 10^2\ s})$ & $[5, 23]$ & $[7, 23]$ & $[9, 23]$ & $[11, 23]$ & $[13, 23]$ & $[15, 23]$ & $[17, 23]$ & $[19, 23]$ & $[21, 23]$ \\
				\hline
			\end{tabular}\label{tab_fault}
		\end{center}
	\end{table*}

The railway network system in this section consists of three trains each of which has three carriages, 
namely, $N=3$ and $M_i=3$ for $i=1,2,3$.
The parameter values of the train model are recorded in Table~\ref{tab_parameter}.
The initial values of the trains are set according to Table~\ref{tab_state}, which satisfy the initial conditions in Theorem~\ref{theorem_con}. 

\begin{figure}[t] 
		\centering 
		\includegraphics[width=0.5\textwidth]{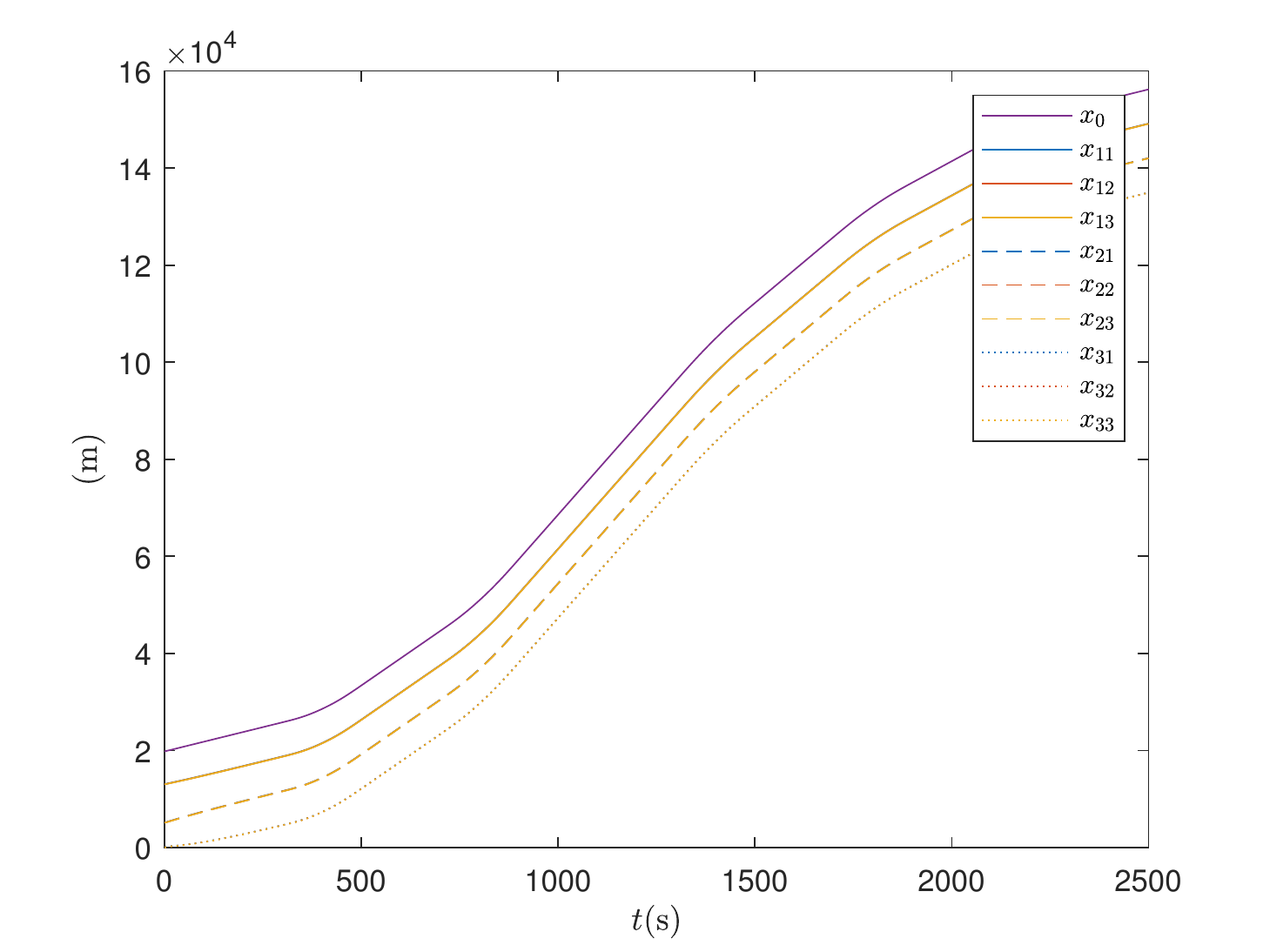} 
		\caption{Profile of the desired position $x_0$ and the positions $x_{ij}$, $i = 1,2,3$, $j = 1,2,3$, of all the carriages.} 
		\label{fig_co_x} 
		\centering 
		\includegraphics[width=0.5\textwidth]{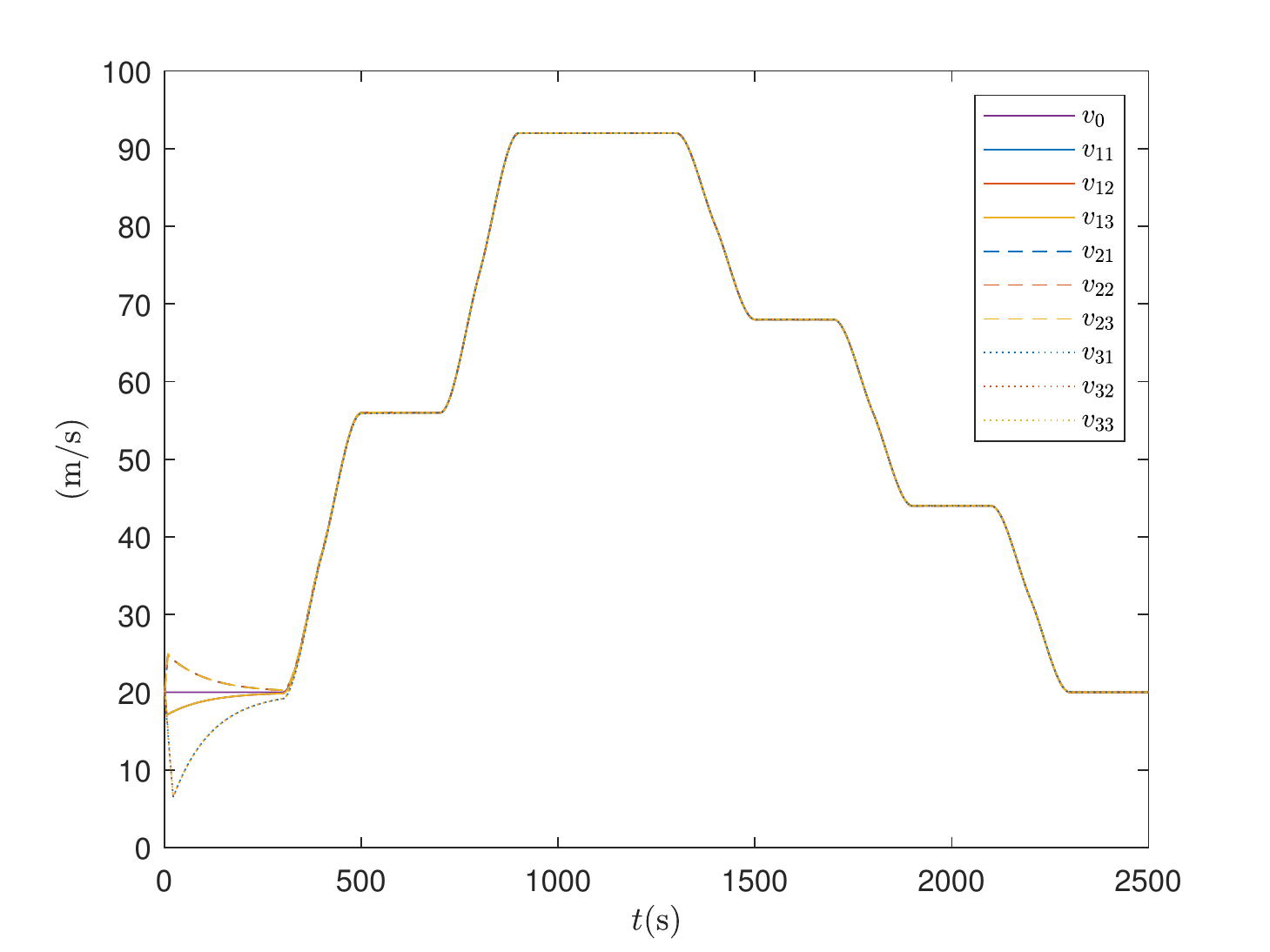} 
		\caption{Profile of the desired velocity $v_0$ and the velocities $v_{ij}$, $i = 1,2,3$, $j = 1,2,3$, of all the carriages.} 
		\label{fig_co_v} 
	\end{figure}

The desired velocity-position profile in \eqref{eq_ref_signal} includes acceleration, deceleration and uniform velocity in multiple periods,
which is shown in Figs.~\ref{fig_co_x} and \ref{fig_co_v}.
 Since the maximum expected velocity is $92{\rm m/s}$, the emergency braking distance is $\gamma_2 = 4702{\rm m}$ (with the acceleration  $-0.9 {\rm m/s^2}$), the service braking distance is $d_s = 7053{\rm m}$ (with the acceleration $-0.6 {\rm m/s^2}$), and the maximum communication distance between vehicles is $\gamma_1 = 9000{\rm m}$. Thus, $\rho_2 = 2351{\rm m}$ and $\rho_1 = 1947{\rm m}$ in \eqref{eq_ds}. The velocity difference between adjacent trains is constrained as $\sigma_2 = \sigma_1 = 50{\rm m/s}$, then $\varrho_2 = 26.5 {\rm m/s} $ and $\varrho_1 = 30.55 {\rm m/s} $ with $\ell_{i1} = 0.01{\rm s^{-1}}$. The nominal value of the position difference is set to $d_p = 26{\rm m}$. 
	
	It is assumed that the actuator of each carriage is invaded by faults, but the action time is not the same. The faults suffered by different actuators are recorded in Table~\ref{tab_fault}, where $t^s_{ij1}$, $t^s_{ij3}$ and $t^e_{ij1}$, $t^e_{ij3}$ are the start time and end time of faults, respectively, and the initial values of faults are set according to Example~\ref{exa_fault}. Constant faults and periodic faults occur randomly and overlap in time periods. In order to highlight the robustness of the proposed scheme, an external disturbance signal is added, which is a Gaussian random number sequence with mean  $0$ and variance  $0.5$. 
	
	The observers are designed as \eqref{eq_observer_1}, and all the eigenvalues of $D_{ij}$, $i = 1, 2, 3$, $j = 1, 2, 3$,  in \eqref{eq_error_all}, are placed at  $-3$ by a proper selection of $k_{ij1}$ and $K_{ij}$. The controllers are designed as \eqref{eq_u_p_ij_1} and \eqref{eq_u_p_ij_3}, in which the parameters are set to $l_{ij1} = l_{ij2} = l_{ij3} = 0.1$, $i = 1,2, 3$, $j = 2,3$, and $\ell_{i1} = 0.01$, $\ell_{i2} = 2.1$, $\ell_{i3} = 4.3$, $\ell_{i4} = 1$, $i = 1,2,3$.
	
\begin{figure}[t] 
		\centering 
		\includegraphics[width=0.5\textwidth]{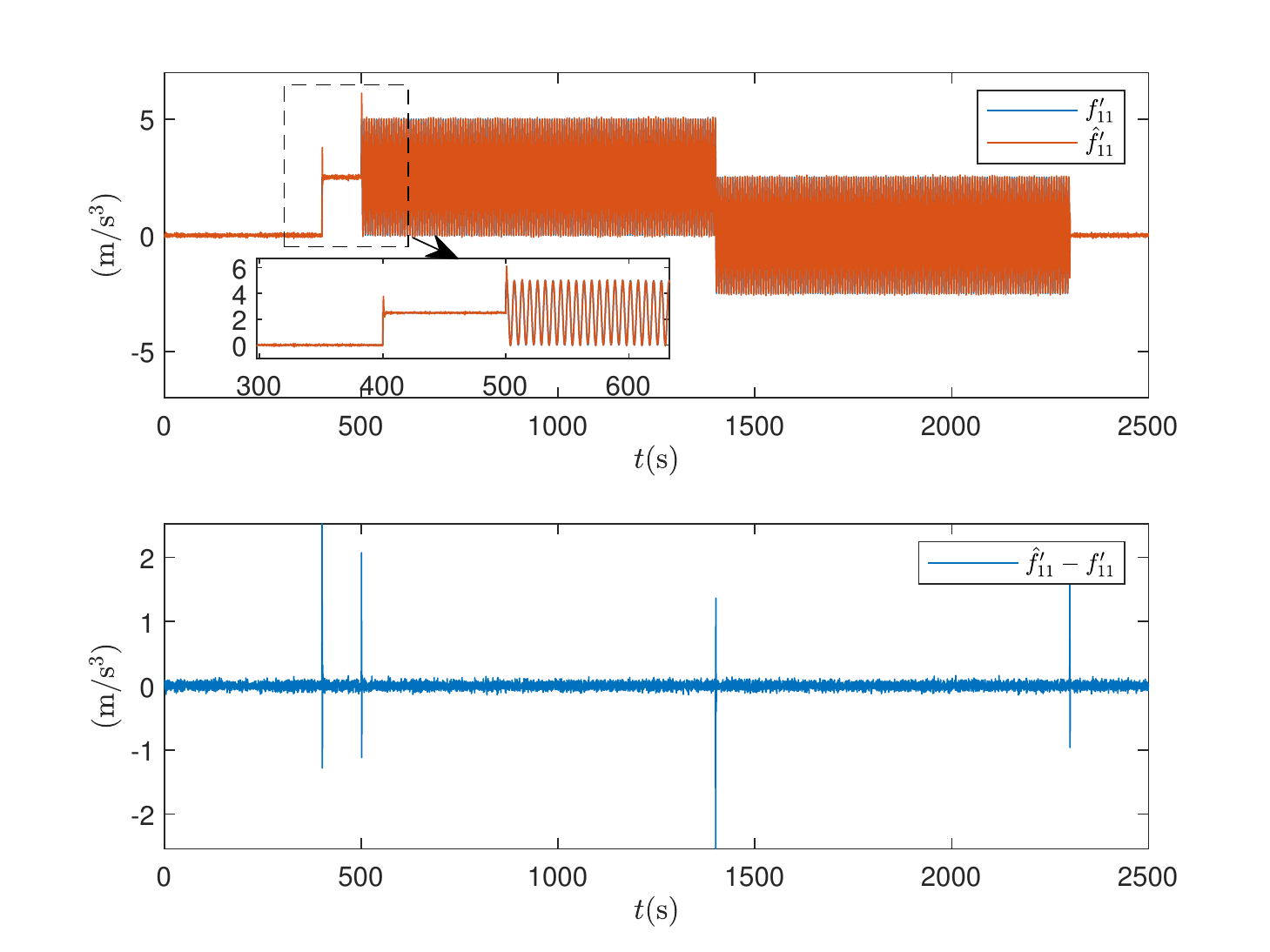} 
		\caption{Top: profile of the fault and its estimation of the carriage $(1,1)$; bottom: profile of the estimation error. } 
		\label{fig_ob_faults_estimation} 
		\centering 
		\includegraphics[width=0.5\textwidth]{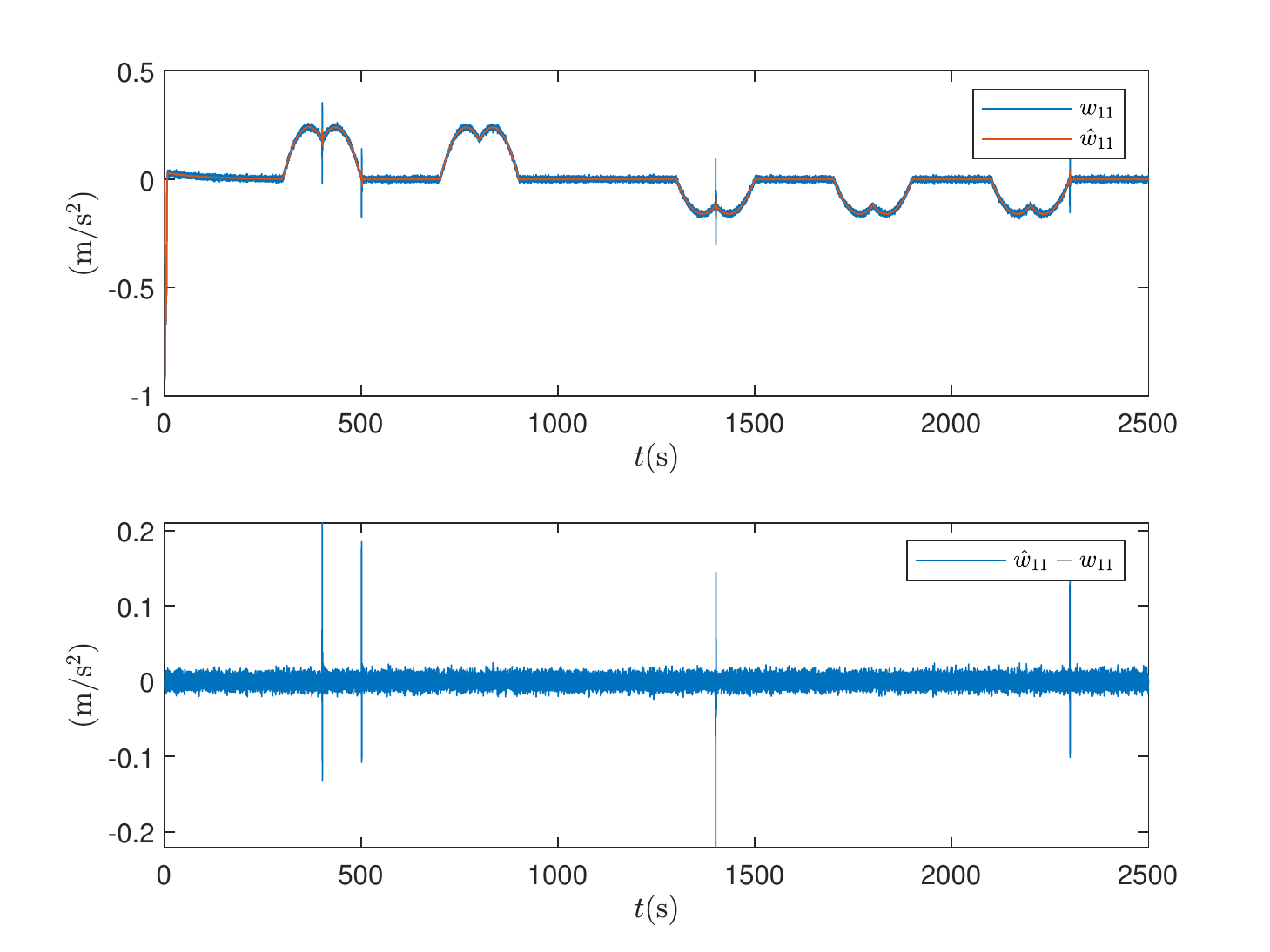} 
		\caption{Top: profile of the acceleration and its estimation of the carriage $(1,1)$; bottom: profile of the estimation error. } 
		\label{fig_ob_w_estimation} 
	\end{figure}

	The results of the numerical experiments are discussed below.  
	Denote $f'_{ij} = C_{ij}f_{ij}$ and $\hat f'_{ij} = C_{ij}\hat f_{ij}$. 
	Estimation of the fault and the acceleration in the carriage $(1,1)$ is shown in Figs.~\ref{fig_ob_faults_estimation} and  \ref{fig_ob_w_estimation}, respectively. It is observed that 
the estimation errors  converge to zero quickly, with residual errors due to the deliberately added Gaussian noise for robustness evaluation. 
The transient fluctuations appear at the moments corresponding to the start time and the end time of the constant and periodic faults. 
 These results demonstrate the superiority of the proposed observer in Theorem~\ref{theo_observer}.
	
\begin{figure}[t] 
		\centering 
		\includegraphics[width=0.5\textwidth]{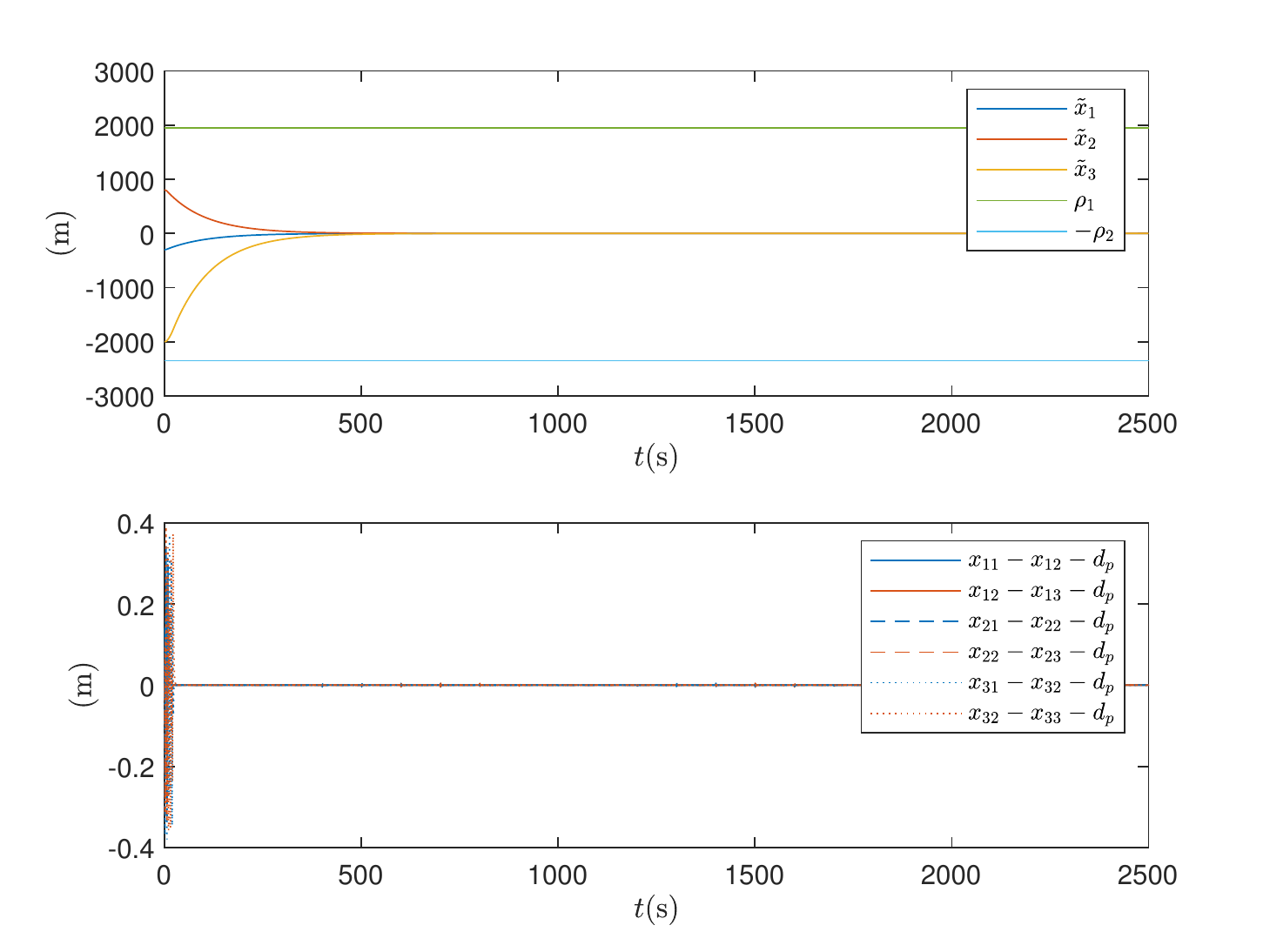} 
		\caption{Top: profile of position difference of adjacent trains;
		bottom: profile of position difference of adjacent carriages.} 
		\label{fig_co_x_d} 
		\centering 
		\includegraphics[width=0.5\textwidth]{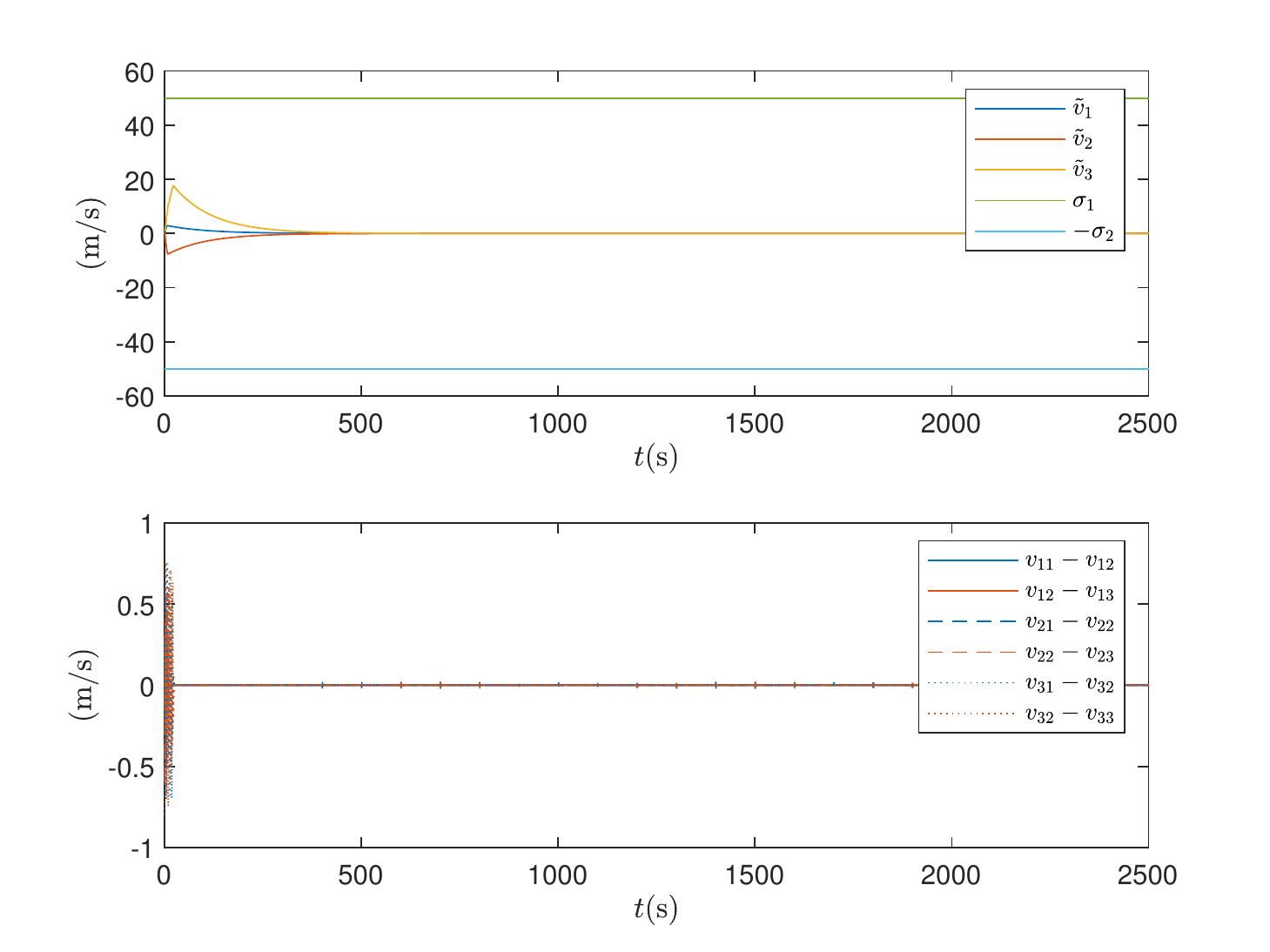} 
		\caption{Top: profile of velocity difference of adjacent trains;
		bottom: profile of velocity difference of adjacent carriages.} 
		\label{fig_co_v_d} 
	\end{figure}

\begin{figure}[t] 
		\centering 
		\includegraphics[width=0.5\textwidth]{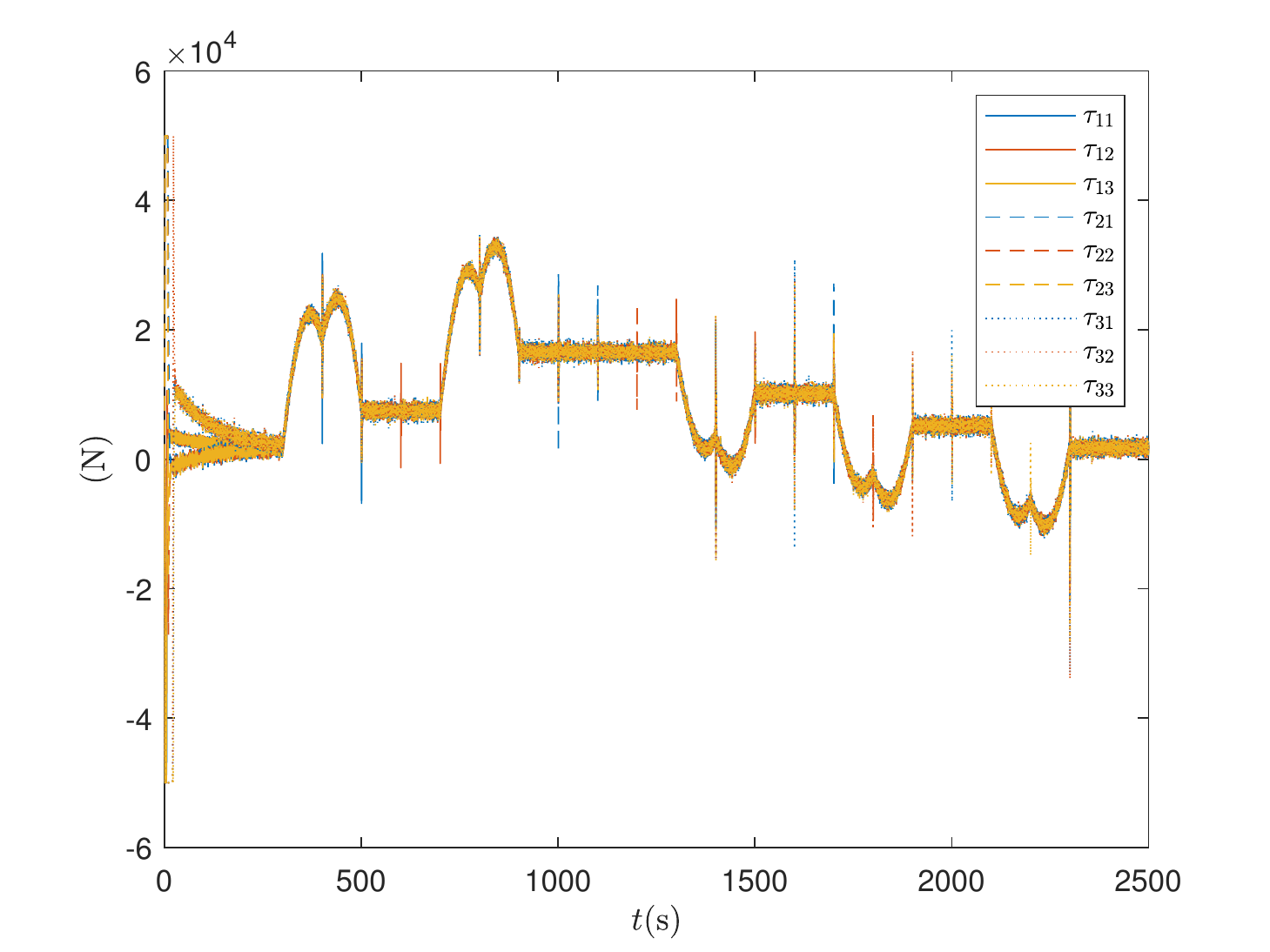} 
		\caption{Profile of the traction/braking forces $\tau_{ij}$, $i = 1,2,3$, $j = 1,2,3$, of all the carriages.} 
		\label{fig_co_tau} 
	\end{figure}

The effectiveness of 	the controller in Theorem~\ref{theorem_con} is demonstrated in Figs.~\ref{fig_co_x_d}  and \ref{fig_co_v_d}.
 The position difference and velocity difference of adjacent trains are within the safety constraint range during the whole operation, and the distance difference eventually converges to the service brake distance $d_s$, and the velocity difference converges to $0$; on the other hand, the position difference between adjacent carriages of every  train   converges to the nominal value $d_p$, and the velocity difference  converges to $0$. 
The overall tracking performance of all the carriages to the reference trajectory in terms of position and velocity is 
shown in  Figs.~\ref{fig_co_x} and \ref{fig_co_v}.   
The required traction/braking forces are plotted in Fig.~\ref{fig_co_tau}.
The transient deviation is observed at the beginning of the simulation due to the initialization of the observers
and also at the start time and the end time of the  faults. 

%
%

%
%
		
\begin{figure}[t] 
		\centering 
		\includegraphics[width=0.5\textwidth]{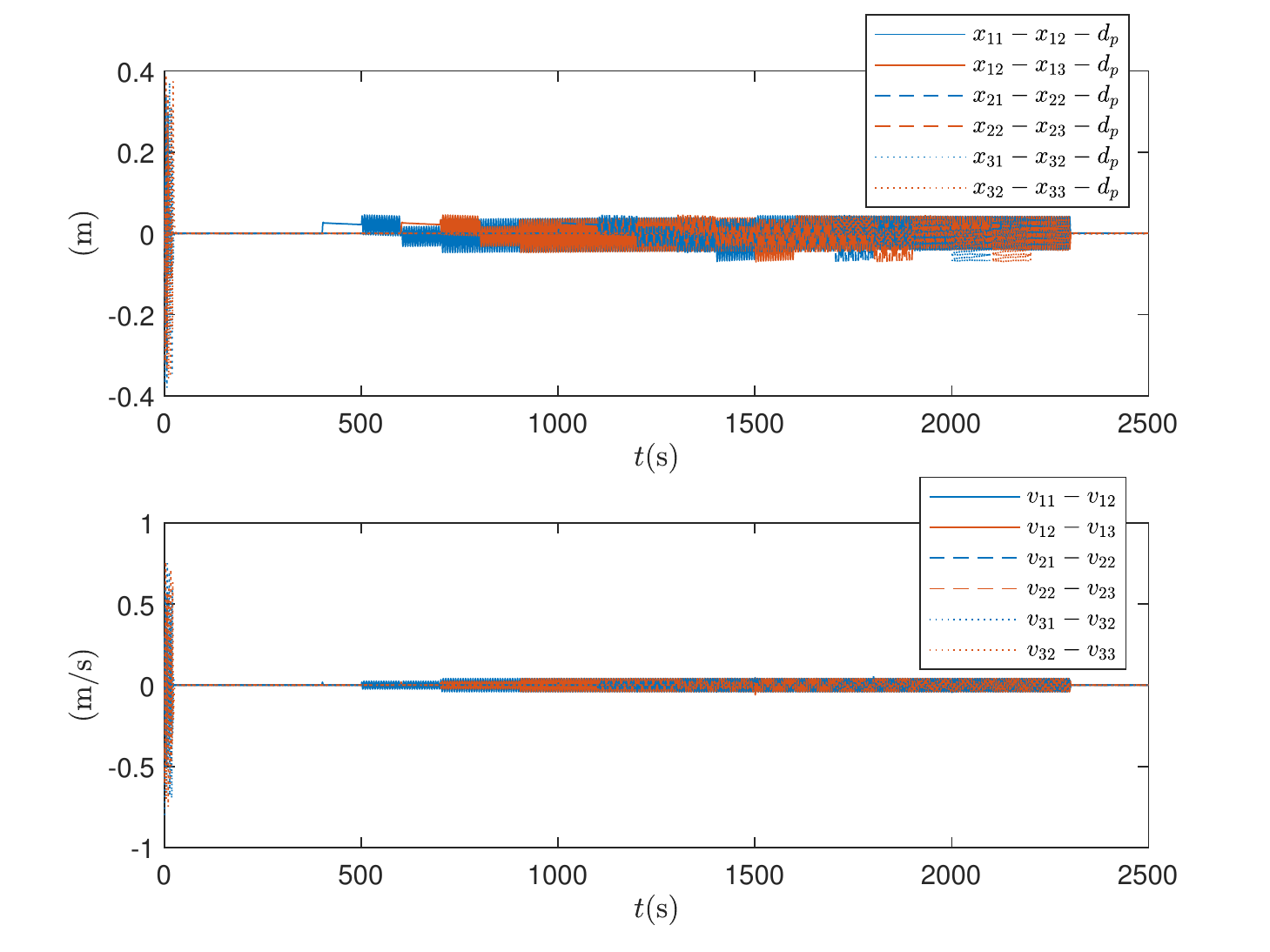} 
		\caption{Profile of position  and velocity differences of adjacent carriages using the observer  in \cite{zhu2015fault}.		 } 
		\label{fig_co_zhu_z_c_x} 
	\end{figure}

Finally, as a comparison experiment, the observer  proposed in \cite{zhu2015fault} is used to estimate faults, 
based on which the same HST model and controller scheme are tested.  
The experimental results are shown in Fig.~\ref{fig_co_zhu_z_c_x}.
Since the observer does not ensure asymptotical estimation, some residual errors appear. 
In particular,  the position and velocity differences of adjacent carriages have more obvious fluctuations  caused by the periodic faults, compared with the results in  Figs.~\ref{fig_co_x_d}  and \ref{fig_co_v_d}.
 This comparison exhibits the superiority of the proposed observer/controller scheme.



	\section{Conclusion} \label{sec_conclusion}
	In this paper, a third-order HST multi-particle model with actuator faults has been 
	established considering the motor dynamics of a real train.  
	The structural information of the faults is also described by the model.  
	Based on this model, an observer for asymptotical estimation of system states and a distributed fault-tolerant controller 
	have been designed to realize  cooperative cruise control under the dual constraints of ensuring the position and velocity differences of adjacent trains in specified ranges  during the whole operation.


	\bibliography{reference}
	\end{document}